\documentclass[useAMS,usenatbib]{mn2e}
\usepackage{graphicx}    
\usepackage{subfigure}
\usepackage[latin1]{inputenc}
\usepackage{amsmath}
\usepackage{amsfonts}
\usepackage{amssymb}

\def \fps@figure{htbp}
\makeatother
\def \aap  {A\&A}

\def \aj  {AJ}
\def \apj  {ApJ}
\def \apjs  {ApJS}
\def \araa  {ARA\&A}

\def \mnras {MNRAS}
\def \nat {Nature}
\def \apjl {ApJL}
\def\pasp{PASP}
\def\apss{Ap\&SS}

\DeclareGraphicsExtensions{.ps,.pdf,.png}
\makeatletter

\begin{document}

\title[Galaxy structure and star formation history]{Linking the Structural Properties of Galaxies and their Star Formation Histories with STAGES.}

\author[Hoyos et al.]{Carlos Hoyos$^{1,2}$\thanks{E-mail: \texttt{carlos.hoyos.f@gmail.com}}, 
Alfonso Arag\'{o}n-Salamanca$^{1}$\thanks{E-mail: \texttt{alfonso.aragon@nottingham.ac.uk}}, 
Meghan E. Gray$^{1}$,
Christian Wolf$^{3}$,
\newauthor
David T. Maltby$^{1}$,
Eric F. Bell$^{4}$,
Asmus B\"ohm$^{5}$,
Shardha Jogee$^{6}$
\\
\footnotemark[0]\\$^{1}$School of Physics and Astronomy, The University of Nottingham, University Park, Nottingham, NG7 2RD, UK
\footnotemark[0]\\$^{2}$Departamento de F\'{\i}sica Te\'orica, Universidad Aut\'onoma de Madrid, Carretera de Colmenar Viejo, 28049 Madrid, Spain
\footnotemark[0]\\$^{3}$RSAA, Mount Stromlo Observatory, The Australian National University, ACT 2611, Australia
\footnotemark[0]\\$^{4}$Department of Astronomy, University of Michigan, 500 Church St., Ann Arbor, MI 48109, USA
\footnotemark[0]\\$^{5}$Institute for Astro- and Particle Physics, University of Innsbruck, Technikerstr. 25/8, 6020 Innsbruck, Austria
\footnotemark[0]\\$^{6}$Department of Astronomy, University of Texas at Austin, 1 University Station C1400 RLM 16.224, Austin, TX 78712-0259, USA}

\date{Accepted ??. Received ??; in original form ??}
\pagerange{\pageref{firstpage}--\pageref{lastpage}} \pubyear{2014}
\maketitle

\label{firstpage}

\begin{abstract}
{We study the links between star formation history and structure for a large mass-selected galaxy sample at $0.05 \leq
z_{\mathrm{phot}} \leq 0.30$. The galaxies inhabit a very broad range of environments, from cluster cores to the field. Using HST images, we quantify their structure following \cite{2012MNRAS.419.2703H}, and divide them into disturbed and undisturbed. We also visually identify mergers. Additionally, we provide a quantitative measure of the degree of disturbance for each galaxy (``roughness''). 
The majority of elliptical and lenticular galaxies have relaxed structure, showing no signs of ongoing star formation. Structurally-disturbed galaxies, which tend to avoid the lowest-density regions, have higher star-formation activity and younger stellar populations than undisturbed systems. Cluster spirals with reduced/quenched star formation have somewhat less disturbed morphologies than spirals with ``normal'' star-formation activity, suggesting that these ``passive'' spirals have started their morphological transformation into S0s. Visually identified mergers and galaxies not identified as mergers but with similar roughness have similar specific star formation rates and stellar ages. The degree of enhanced star formation is thus linked to the degree of structural disturbance, regardless of whether it is caused by major mergers or not. This suggests that merging galaxies are not special in terms of their higher-than-normal star-formation activity. Any physical process that produces ``roughness'', or regions of enhanced luminosity density, will increase the star-formation activity in a galaxy with similar efficiency. An alternative explanation is that star formation episodes increase the galaxies' roughness
similarly, regardless of whether they are merger-induced or not.}
\end{abstract}


\section{Introduction}
\label{sec:Introduccion}

\subsection{Background}
\label{sec:Background}

Collisions and interactions between galaxies can significantly change their morphologies and star-formation histories. 
In this paper we try to link these two aspects of galaxy evolution.
Mergers can change the number density of galaxies and contribute to 
the shape and colour dependence of the luminosity function \citep{2003A&A...401...73W,2004ApJ...608..752B,2007ApJ...665..265F}.
They are therefore key ingredients in hierarchical galaxy evolution models. Furthermore, it
has been argued that merger events are expected
to drive the formation and evolution of massive early-type galaxies \citep{2007MNRAS.375....2D,2010ApJ...719..844R,2012arXiv1202.4674L,2011MNRAS.415.3903T,2012ApJ...747...34B}, 
although their impact for early-type galaxies of intermediate mass could be lower
\citep{2010ApJ...710.1170L,2010A&A...518A..20L}.

One way to identify galaxy mergers is to measure the galaxies' structural properties in optical images
\citep[see, e.g.,][]{2003ApJS..147....1C,2004AJ....128..163L,2009ApJ...697.1971J}. Alternative
merger detection techniques are based on the identification of kinematical and spatial close pairs 
\citep{2000ApJ...536..153P,2009ApJ...692..511W}. The morphological techniques are based on the fact that the objects
involved in a merger episode will be gravitationally disturbed and appear to be either highly asymmetric or shredded on deep
images. \citet{2012MNRAS.419.2703H} introduced a new method to identify galaxy mergers based
on the structural properties of the residual images of galaxies after subtracting a smooth model. This diagnostic
quantifies the deviation of the surface brightness profiles of galaxies from a S\'ersic model and it was optimized 
to detect almost all minor mergers.

{It is important to point out that none of the methods which employ structural diagnostics 
can produce a clean merger sample. This is particularly true for minor mergers,
where the contamination by non-mergers can be rather high (up to 70\%; \citealt{2012MNRAS.419.2703H}). 
These contaminants can be late-type galaxies with prominent and localized star-forming 
regions, or objects disturbed by tidal effects.
Thus, it is important to keep in mind that many galaxies with disturbed morphologies are not necessarily mergers. }

There are many factors that influence the structural nature of a merger remnant {and thus our ability to identify mergers.} 
These include the mass ratio 
between the colliding galaxies, the morphological type of the merging systems, and the amount of gas present in them.
Other relevant parameters include the orbital elements of the collision which can, for instance,
determine whether or not a tidal dwarf galaxy is formed after the merger \citep{2011arXiv1101.4834D}. The prototypical
result of these studies is that the collision between two equally-massive large disk galaxies generally results in an elliptical galaxy
\citep{1977egsp.conf..401T,1991ApJ...371...92S,1991ApJ...370L..65B,1982ApJ...252..455S}.
It has also been found that there is a mass ratio interval between \texttt{3:1} and \texttt{4.5:1} for which the
remnant of a merger between two disk galaxies can be either an elliptical galaxy, a galaxy with a disturbed
disk or a hybrid  system with disk morphology but with the kinematics of an elliptical galaxy
\citep{2004A&A...418L..27B,2005MNRAS.357..753G,2005A&A...437...69B}. However, \citet{2006MNRAS.369..625N} showed 
that it is often problematic to assess the intrinsic structural nature of a merger remnant observationally.
Moreover, multiple minor mergers can transform a disk galaxy into an early
type system, as shown by \citet{2007A&A...476.1179B}. This transformation is most clearly seen in the evolution of the morphological S\'ersic
index and the kinematic $V_{\mathrm{rot}}/\sigma$ ratio of the final system after each minor merger event.
These works highlight that there must be a relation between the conditions in which a merger takes place and the 
structural properties of the resulting galaxy.

Gas also plays an important role in determining the nature of merger remnants, and its importance
is far greater than its relative mass with respect to the stellar component. For instance, \citet{2010MNRAS.404..590L}
show that gas-rich mergers appear to be more disturbed and asymmetric than dry mergers, and they remain so for longer.
The reason for the enhanced importance of the gaseous component of galaxies
lies with the way it reacts to an encounter \citep{1951ApJ...113..413S}. Gas can 
collide and compress while readjusting to the rapidly-changing gravitational potential of the merger. It 
can also undergo efficient radiative cooling and, as a result, be converted into new young stars in a 
central starburst episode or feed an active galactic nucleus. Gas thus plays a key role in the origin of the relation between the
mass of the central black hole and the velocity dispersion of the stars in galaxies
\citep{1996ApJ...464..641M,1996ApJ...471..115B,2005Natur.433..604D,2005ApJ...630..705H,2008A&A...492...31D,2009ApJ...707L.184J}.
Gas-rich mergers are also thought to be very important for the evolution of the stellar populations
found in early-type galaxies. \citet{1999Ap&SS.267..299S} shows that several distinct stellar populations can be formed
during a merger event through a series of separate starbursts taking place during the final mixing phase of the merger.

{Taking all these factors into account, it is clear that unambiguously identifying mergers is not an easy task, particularly in the
absence of kinematic information revealing disorder in the velocity field. Furthermore, minor mergers, in particular gas-poor ones, are much more difficult to identify than major mergers. It is clear therefore that imaging data alone cannot unambiguously distinguish between mergers and non-mergers in all cases. However, when visually identifying mergers it is possible to draw on the experience the classifiers have built by examining the images generated by mergers simulations and the expertise of previous classifiers. Asymmetry or ``roughness'' is not enough to classify a galaxy as a visual merger. Signs such as tidal tails, double/multiple nuclei, clear bridges of material connecting different galaxy components, etcetera, are required. Even though visual image classification cannot be completely objective and unequivocal,  the classifiers' experience can be very valuable and provide useful information, as we will show in this paper.}

The impact of mergers on the star-formation histories of galaxies has been
discussed in detail in the literature. The work presented in
\citet{2003A&A...405...31B} finds a moderate effect of mergers on the global
SFR in galaxies at $z\sim0$, although they also find that most of the
merger-induced contribution to the SFR of galaxies is localized at the very
centres of the interacting galaxies. Relatively recent works include those of
\citet{2009ApJ...697.1971J} and \citet{2009ApJ...704..324R}, which use data
from the GEMS\footnote{See \texttt{http://www.mpia.de/GEMS/gems.htm}} 
\citep{2004ApJS..152..163R} and \textsc{COMBO-17}
\citep{2003A&A...401...73W,2004A&A...421..913W,2008A&A...492..933W,2004ApJ...608..752B} 
surveys respectively. These
studies conclude that in massive galaxies the star formation rate is moderately enhanced in
interacting and merging systems compared to non-interacting systems of similar
masses over the $0.2\leq z \leq 0.8$ redshift interval
\citep{2005ApJ...630..771W,2005ApJ...625...23B}. Moreover, the work
shown in \citet{2009ApJ...705.1433H}, based on the same dataset
we analyse here, shows that the star-formation rate in mergers
is a factor 1.5 -- 2.0 times higher in mergers and interacting galaxies
than in non-interacting systems. However, since mergers only constitute a small
(5\% -- 10\%) fraction of the total population of cluster galaxies, their
contribution to the \emph{global} star formation  in the A901/902
multiple cluster environment is only between 10\% and 15\%.

Additional mechanisms which could affect 
the evolution of cluster galaxies and thus alter their
morphologies and star-formation histories include:
ram-pressure stripping of the gaseous component as the galaxy travels through the densest regions of the Intra Cluster Medium 
\citep{1980ApJ...237..692L,1972ApJ...176....1G}; gas compression \citep{1998ApJ...509..587F}; thermal evaporation 
\citep{2007MNRAS.382.1481N}; and
frequent high-speed encounters between galaxies \citep[galaxy harassment,][]{1998ApJ...495..139M}. 
All these phenomena leave their own imprints on both the morphologies and star formation histories
of galaxies. For instance, \citet{2004ApJ...613..866K} observe disk truncation of the H{\small\rm II} line emission in Virgo cluster galaxies. However, the regimes in which each of these phenomena act and the time scales
of their effects need not be the same \citep{2006PASP..118..517B}, although all of them tend to transform a blue star forming galaxy
into a red passively evolving one.

The morphology--density relation \citep{1977ApJ...215..401M,1980ApJ...236..351D} provides clear evidence of
the connection between the structure of galaxies and the environment where
they live. Early-type galaxies are 
preferentially found in high-density environments, whereas late-type galaxies
are preferentially located in low-density environments. This relation 
has been in place since $z \sim 1$ \citep{2005ApJ...620...78S,2008ApJ...684..888P}. 
Star formation is also known to be suppressed in high-density environments
\citep{1998ApJ...499..589H,2002MNRAS.334..673L,2004MNRAS.347L..73G,2006ApJ...642..188P}, even 
after factoring out the underlying morphology--density relation.

A basic prediction of merger models is that substructure in either late mergers or 
recent merger remnants should be associated with signatures of a young stellar population. This has already been 
successfully tested by \citet{1992AJ....104.1039S} and \citet{2010ApJ...724..694G} using ground-based imaging data 
for representative samples of local galaxies. Here, we plan to further these studies using a well-defined galaxy sample
with good space-based imaging. On the other hand, optically-passive (sometimes called \textit{anemic}), 
low-star-formation spirals should show smoother morphologies than other spiral systems of similar masses. 
These systems, first identified by \citet{1976ApJ...206..883V} and more recently
by \citet{1999ApJ...518..576P}, are spiral galaxies with smooth arms and no evident signs of active star forming regions.
These sources were shown in \citet{2009MNRAS.393.1302W} to be massive ($10.0< \log M / M_{\odot} < 11.0$) spiral 
galaxies that probably have been incorporated into the cluster medium in
recent times, a view that is strengthened by \citet{2013A&A...549A.142B}. Their star formation rates have been significantly
reduced and they are thought to have just started a slow morphological transformation into smoother S0 galaxies.
These objects are mostly the so-called ``dusty red'' systems described and characterized in \citet{2005A&A...443..435W}. Optically-passive spirals are then systems in which the star formation history has already started to evolve amidst the cluster environment but whose
large-scale structural properties are still largely unchanged, as revealed by their spiral morphologies. We will call these galaxies ``red spirals'' in what follows.

\subsection{Scope of this paper}
\label{sec:Scope}
Taking all of this into consideration, it is clear that different galaxy transformation processes will leave different imprints
on the structure \emph{and} star-formation history of the galaxies. It is also clear that any study trying
to link the morphological disturbances seen in galaxies with their star formation histories would benefit from 
considering  galaxies in a broad range of environments, including the field, groups, clusters, and the infall regions. 

In this paper we further explore the relation between the star formation
histories of galaxies and their degree of morphological disturbance by
expanding the work presented in \citet{2012MNRAS.419.2703H}. In that
paper we used the \textsc{STAGES} 
\citep[Space Telescope A901/902 Galaxy Evolution Survey,][]{2009MNRAS.393.1275G} \textit{Hubble Space Telescope} (HST) 
\textit{Advanced Camera for Surveys} (ACS) F606W images to develop a structural merger diagnostic 
using a complete sample of both cluster and field galaxies with a well-defined
mass limit and reliable visual morphological classifications.
Thus, one goal of the current work is to assess whether the morphological indicator introduced in \citet{2012MNRAS.419.2703H}
correlates with (or provides  information about) the star formation and/or merging
histories of galaxies. In particular, we here
place the spotlight on (i) confirmed visual mergers, (ii) disturbed objects that are however not {visually classified 
as mergers}\footnote{{As mentioned in Section~\ref{sec:Background}, this class probably contains a difficult-to-quantify 
fraction of minor and/or late-stage mergers that defy visual identification, and any interpretation of the results must take this caveat into account.}}, (iii) relaxed, 
non-disturbed objects and (iv) red spiral galaxies. We use the
luminosity-averaged ages, star formation rates (SFR) and morphological
disturbances of these different galaxy sets in order to explore the
link between star formation histories and morphology and, in particular, whether this 
relationship depends on the physical origin (i.e, merger induced or internally generated) of the observed
morphological disturbances.

Objects are considered to be ``disturbed'' if they lie in the merger side of the structural diagnostic defined in 
\citet{2012MNRAS.419.2703H}, and galaxies are said to be ``relaxed'' otherwise. In addition, we define the sub-sample of visual mergers as the systems identified by eye as mergers in \citet{2012MNRAS.419.2703H}.
Besides the obvious interest in comparing the properties of the stellar population histories found in actual
mergers to those found in the relaxed systems, the properties of the disturbed objects that are visually classified 
as non-mergers are interesting because they can help decide whether the enhanced star formation 
found in mergers can also be present in non-merging disturbed galaxies. {However, some caution is needed here. In order for this to work in all cases, it would be necessary to have a diagnostic that can cleanly identify all the non-mergers. The visual diagnostics that we use may sometimes fail to separate minor mergers from clumpy star-forming disks. When the star-forming clumps are clearly part of the spiral arm structure, without disturbing the overall symmetry of the galaxy, is reasonable to assume that they are not the result of minor mergers; however, when these clumps are not clearly part of the spiral arms, it is not possible to ascertain, from the images alone, whether they are due to mergers or to other causes. Additional information, such as kinematics, would be very useful here.}

Another goal of this work is to assess whether the star-formation histories of {the disturbed} non-merging 
systems are comparable with those of similarly-disturbed \emph{merging} galaxies {(see below for a quantitative definition
of disturbance)}.
Significant structural perturbations, possibly leading to star formation
events affecting the optical appearance of non-merging systems would need to be explained via internal or perhaps
environmental mechanisms, and would help understand
better the nature of the galaxies identified with our structural merger
diagnostic. It is also possible to consider that a disturbed surface
brightness profile can have various origins, ranging from an actual merger to
a simple nuclear starburst or a simple supernova event. This study can help
ascertain whether all these possibilities leave specific imprints on the
morphologies of their host galaxies.
Finally, the set of cluster red spirals will be used to explore their degree
of morphological transformation with respect
to other spirals in the cluster environment. The comparison of red
spirals with other spirals is more important in the cluster environment since
red spirals are fairly scarce in the field.

We stress that the work we present here is purely phenomenological. It may be possible to use high-resolution
hydrodynamical simulations of galaxy mergers and normal star-forming systems
in order to tackle these complicated questions, assessing, for instance, the typical temporal
and/or spatial scales of star formation episodes in mergers and in isolated
galaxies, or other structural traits. However, this theoretical work is
very difficult to do since it is not clear what to look for and the detailed physics is poorly constrained.  
Indeed, it would be
an enormous computational challenge to simulate a complex and coupled system of cold and hot gas, stars, dust and dark 
matter in a dynamic cluster environment. Therefore, we will focus on the conclusions that can be drawn from the observations alone.

Section~\ref{sec:DatosMuestra} presents the observational data and galaxy sample.
Section~\ref{sec:Binario} shows the basic differences between the star formation histories of 
smooth, disturbed, and merging galaxies.
In section~\ref{sec:Continuo} we present the search for relations between the
degree of morphological disturbance in galaxies and their star formation
histories.
Finally, section~\ref{sec:Conclusiones} summarizes the results of this work.
Throughout this paper, a concordance cosmology with $\Omega_{m}=0.3$, $\Omega_{\Lambda}=0.7$ and 
$H_{0}=70\,\mathrm{km}\,\mathrm{s}^{-1}\mathrm{Mpc}^{-1}$ is used.

\section{Data and Sample}
\label{sec:DatosMuestra}

\subsection{Data}
\label{subsec:Datos}

In this work, we make use of the \textsc{STAGES}\footnote{\texttt{http://www.nottingham.ac.uk/astronomy/stages}} dataset \citep{2009MNRAS.393.1275G}. One of the primary goals of \textsc{STAGES} is to study the impact of environment on galaxy evolution, using
the multiple cluster system Abell 901/902 located at $z \simeq 0.165$. This complex comprises different
environments with galaxy densities reaching $n\sim 1000\,\mathrm{Mpc}^{-3}$ (see \citealt{2009ApJ...705.1433H} 
for details). 
\textsc{STAGES} makes use of a wide array of data, including, 
but not limited to, deep HST/ACS imaging, \textit{Spitzer} Infrared photometry and ground based optical observations.

The \textsc{STAGES} HST/ACS observations cover almost $30\arcmin \times 30\arcmin$ in the F606W filter, and were
reduced with a target plate scale of 0.03 arcsec per pixel. The Point Spread Function (PSF)
Full-Width-Half-Maximum (FWHM) is 3.12 pixels or $0.094\arcsec$. At $z=0.165$, 1 arcsec subtends 2.83 kpc, 
giving a spatial resolution of 0.3 kpc. The average exposure time is $\simeq 2\,$ks and
the point-source completeness limit of these images is $F_{\mathrm{606W,AB}}=28.5$.
These observations can provide reliable structural parameters up to an apparent magnitude
between $m_{\rm F606W,AB}=24.0$ and $m_{\mathrm{F606W,AB}}=25.0$, as shown in \citet{2009MNRAS.393.1275G} and \citet{2012MNRAS.419.2703H}. Visual inspection of these frames by multiple independent classifiers provides a robust morphological catalogue, which will be presented in
\citet{StagesMorfoCat}.

The STAGES HST imaging dataset is complemented with \textsc{COMBO-17}
\citep{2003A&A...401...73W,2004A&A...421..913W,2008A&A...492..933W} good-quality optical images 
gathered with the Wide Field Imager (WFI) at the 
MPG/ESO 2.2-m telescope. Additionally, 24$\micron$ observations were obtained with the MIPS instrument on \textit{Spitzer}
\citep{,2009ApJ...690.1883G}. 
The \textsc{COMBO-17} observations consist of optical images in five broad-band filters (UBVRI) and 12 medium-band
filters whose wavelengths range from 3500 to 9300$\,$\AA, covering $34\arcmin\times 33\arcmin$ with a plate scale of 0.238 arcsec
per pixel. The observations are particularly deep in the R band, with an exposure time of 20$\,$ks.
The MIPS data have a plate scale of 1.25 arcsec per pixel and a PSF FWHM of 6 arcsec, and are 80\% complete at a flux density of
97.0 $\mu\mathrm{Jy}$.

The \textsc{COMBO-17} observations yielded very accurate 
photometric redshifts with uncertainties of $\sigma(cz)\simeq 1500$\,$\mathrm{km}\,\mathrm{s}^{-1}$. This corresponds to 3\% distance
errors at the cluster distance for galaxies 
with $R_\mathrm{Vega}<20.0$ \citep[see][]{2009MNRAS.393.1275G}, while the corresponding uncertainty for
$R_\mathrm{Vega}=23.5$\footnote{For the sake of brevity we will omit the subindex ``Vega'' hereafter, but
it applies to all \textsc{COMBO-17} magnitudes and colours discussed in this
paper.} galaxies is $\sigma(cz)\simeq 6000\,\mathrm{km}\,\mathrm{s}^{-1}$, or a 12\%
distance error.
Note that the \textsc{COMBO-17} colours mostly probe the central regions of
galaxies due to the use of relatively small apertures \citep{2009MNRAS.393.1302W}.
It is also important to point out that \textsc{COMBO-17} photometry is much
shallower than the HST \textsc{STAGES }observations.
The available data were also used to estimate Star Formation Rates (SFR), 
photometric stellar masses ($M_{\ast}$), integrated central colours and rest-frame SEDs.

SFRs were obtained using a combination of MIPS observations and \textsc{COMBO-17} rest-frame synthetic 2800 \AA\ luminosities,
calibrated with a set of \textsc{PEGASE} \citep{1997A&A...326..950F} models. In the framework of the models used, the derived SFRs are averages over the last $\sim100\,$Myr assuming a constant SFR.
The \textsc{STAGES} data catalog \citep{2009MNRAS.393.1275G} lists three distinct determinations of the SFR.
The most informative one is called \textit{sfr\_det}, which combines actual measurements of the \textit{Spitzer} $24\micron$ flux 
with the synthetic 2800 \AA\ luminosities. This measurement is used whenever possible. In the absence of a reliable $24\micron$ flux
detection, it is possible to obtain a lower (listed as \textit{sfr\_lo}) limit and an upper (given as \textit{sfr\_hi}) limit to the SFR, and both values
are averaged together. If the resulting averaged value is less than $0.14\,M_{\odot}\,\mathrm{yr}^{-1}$, it is considered an \textit{upper} limit. This limit to the SFR of $0.14\,M_{\odot}\,\mathrm{yr}^{-1}$  is simply the \textit{Spitzer} $24 \micron$
detection limit at $z=0.165$, and for field galaxies this
limiting value is modified taking into account their luminosity-distance.
In the very few cases where the average of the two SFR determinations exceeds the $0.14\,M_{\odot}\,\mathrm{yr}^{-1}$ threshold, this mean value is adopted as the actual value.
Finally, in the absence of \textit{Spitzer} observations\footnote{Not the same as a non-measurement. This is mainly due to spatial coverage issues or to the presence of a bright
star in the middle of the field which makes the \textit{Spitzer} measurements unreliable.} the SFR given by \textit{sfr\_lo} is taken as a lower limit.

Stellar masses were estimated following the work presented in 
\citet{2006A&A...453..869B} using a template library built again using the \textsc{PEGASE} models.
See \citet{2009ApJ...690.1883G} for further details on the models used for the SFR and photometric mass estimates.

SEDs were classified into three different bins describing old red galaxies, ``dusty red'' galaxies  (called red spirals in this paper), and blue cloud 
objects according to their location in the $E_{B-V}$ \textsl{vs.} $(U-V)_{\mathrm{Rest\ Frame}}$ plane. Red spirals were observationally defined in
\citet{2005A&A...443..435W} as having
$(U-V)_{\mathrm{Rest\ Frame}}>(U-V)_{\mathrm{CMR}}-0.25$ and
$E_{B-V}>0.1$. The $(U-V)_{\mathrm{CMR}}$ term is the colour-magnitude relation, defined as
$(U-V)_{\mathrm{CMR}}=1.48-0.4 \times z-0.08 \times (M_{V}-5\times \log h
+20)$. In the previous expression, $h$ is the reduced Hubble-Lema\^itre constant ($H_{0}=100 \times h\ \mathrm{km}\,\mathrm{s}^{-1} \mathrm{Mpc}^{-1}$).

The \textsc{COMBO-17} data were also used to derive aperture stellar mass surface 
densities ($\log \Sigma_{\mathrm{300\ kpc}}^{M}(>10^{9}M_{\odot})$) for cluster galaxies. This is the integrated stellar mass density of galaxies more massive than $10^{9}M_{\odot}$ in apertures with a $300\,$kpc radius (1\farcm75 at $z=0.165$). As shown in \citet{2009MNRAS.393.1302W}, stellar mass densities are more robust against different selection criteria than simple number densities as long 
as the massive $L^{\ast}$ galaxies are included in the sample. These stellar
mass density data were used in order to explore (i) whether the long term star
formation histories of cluster galaxies is a function of the specific
environment and (ii) whether the relaxed, disturbed, and visual merger
cluster galaxies actually exist in the same environmental conditions.

\begin{figure*}
\centering{
\subfigure{
\centering{\includegraphics[scale=0.45]{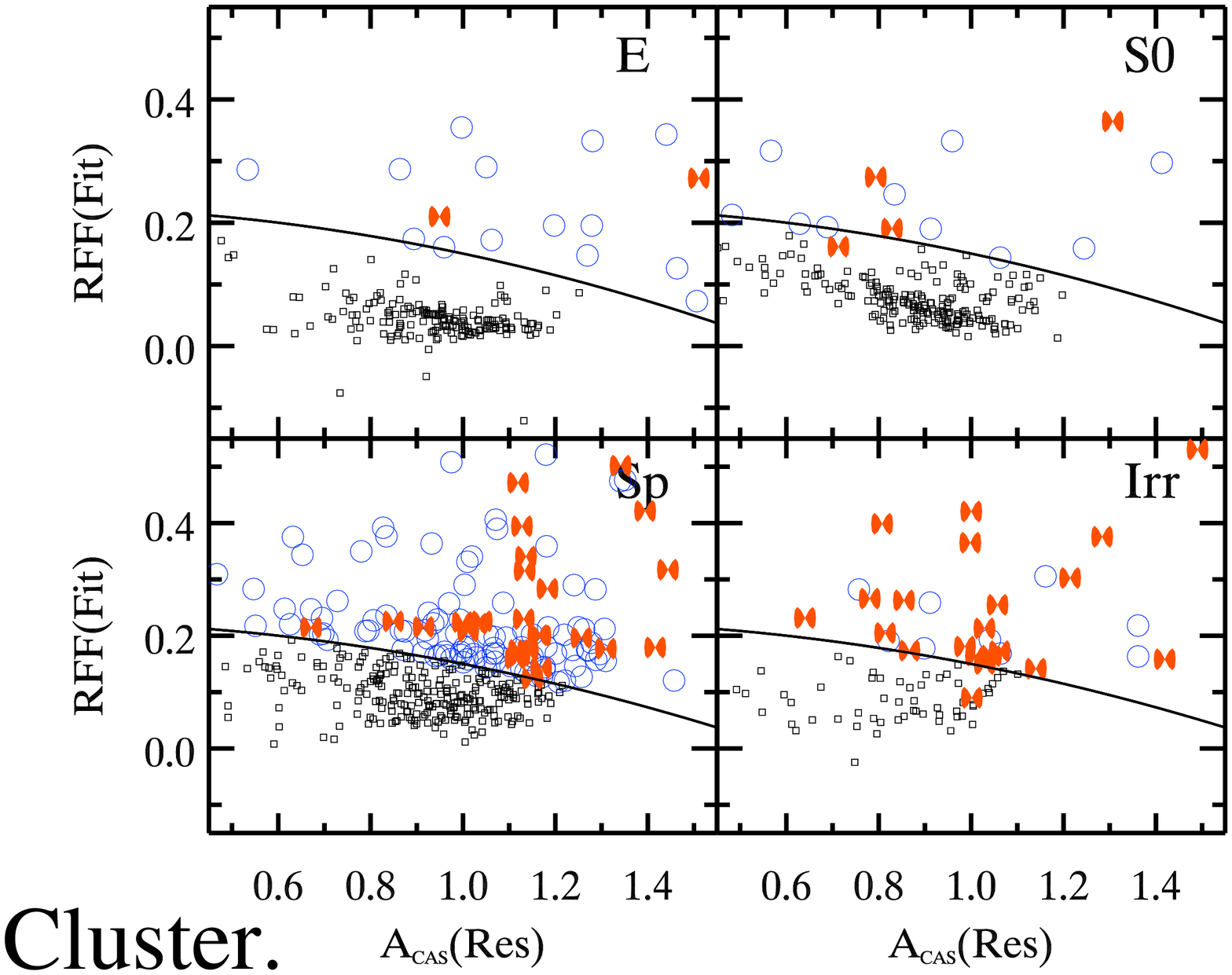}}}
\subfigure{
\centering{\includegraphics[scale=0.45]{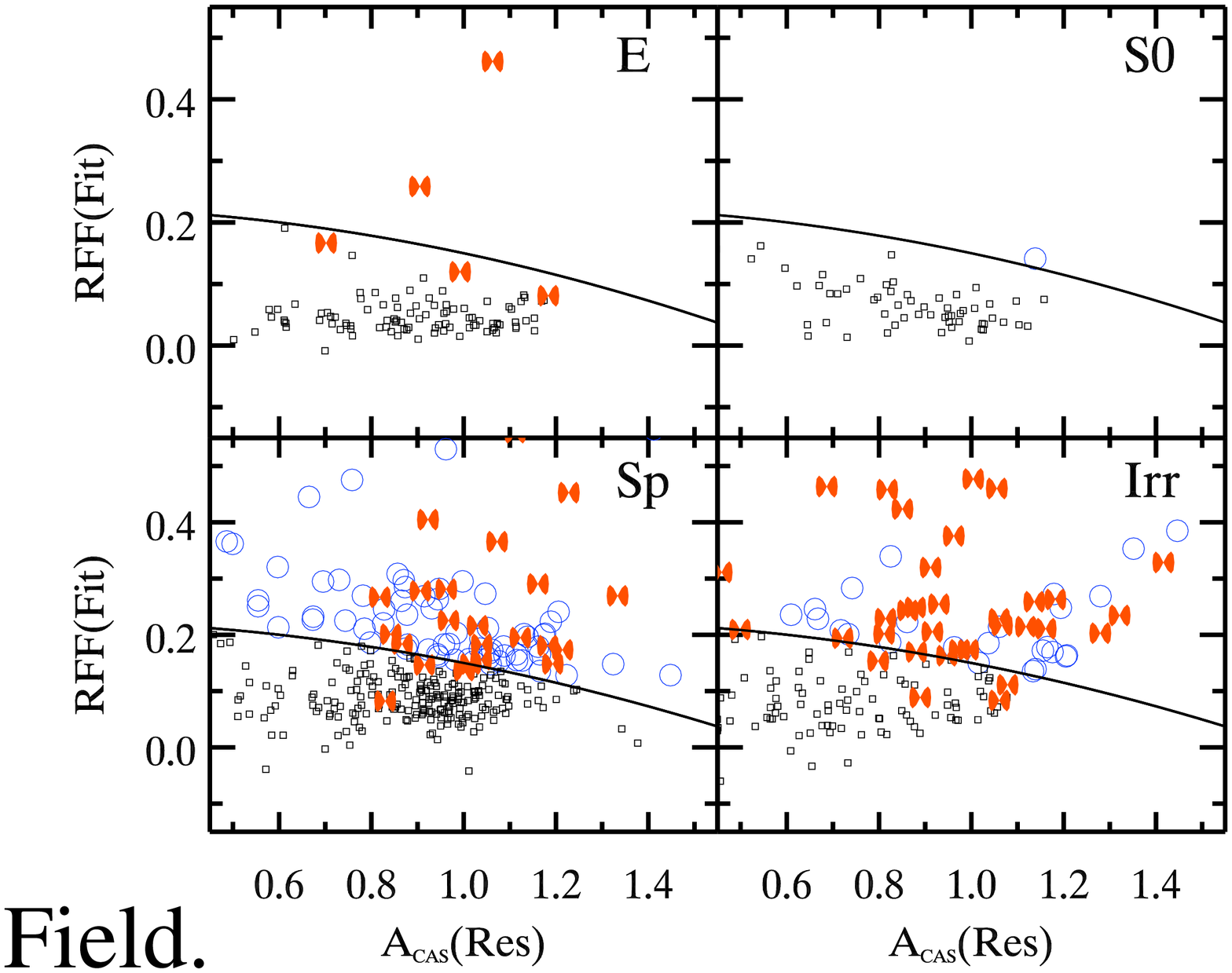}}}
\caption{Subsample definitions used in this work. This plot shows the \textit{RFF}--$A_{\mathrm{CAS}}(\mathrm{Res})$ structural 
parameter space as a function of environment and morphological type. The top panel shows galaxies in the cluster environment and the
bottom panel presents galaxies in the field environment. Each panel includes galaxies of all four morphological types considered in 
this paper, E (top left), S0 (top right), Sp (bottom left), and Irr (bottom right) as indicated.
Relaxed systems are represented as small black hollow squares ({\tiny$\square$}),
disturbed objects {not classified as mergers} 
are depicted as blue circles ($\bigcirc$), and visually-classified mergers are shown as red filled
infinity symbols ($\infty$). Note that the vast majority of {visual} mergers are ``disturbed'' objects, 
but many ``disturbed'' objects are not visually classified as mergers, {although some of them may be minor and/or advanced mergers}. The black line is the boundary that separates objects considered to be relaxed from galaxies considered to be disturbed for the purposes of this work. This figure is thus a graphical representation 
of the main subsamples used in this paper.  
\label{fig:binaryMergers0}}
}
\end{figure*}

\begin{table*}

\begin{center}
\caption{Fractions of relaxed and disturbed non-mergers, and visual mergers for each
  morphological type and environment. Note that only galaxies not classified
  as visual mergers are included into the relaxed and disturbed categories.\label{tab:Fracciones1}}
\begin{tabular}{lccc}
\hline
{Type/Environment} & {Relaxed} & {Disturbed} & {Visual Mergers}\\
{} & {(non-mergers)} & {(non-mergers)} & {}\\
\hline
E & & & \\  
\qquad Cluster & 0.91$\pm$0.02 &	0.08$\pm$0.02 &	0.01$\pm$0.01  \\			
\qquad Field   &   0.94$\pm$0.02 &	0.01$\pm$0.01 &	0.05$\pm$0.02  \\ 
\hline			
S0 & & & \\
\qquad Cluster & 0.93$\pm$0.02 &	0.05$\pm$0.02 &	0.02$\pm$0.01 \\
\qquad Field   &   0.97$\pm$0.02 &	0.03$\pm$0.02 &	0.00$\pm$0.00 \\ 
\hline	
Sp & & & \\
\qquad Cluster & 0.65$\pm$0.02 &	0.27$\pm$0.02 &	0.08$\pm$0.01 \\			
\qquad Field   &   0.70$\pm$0.03 &	0.22$\pm$0.02 &	0.08$\pm$0.02 \\ 
\hline			
Irr & & & \\
\qquad Cluster & 0.58$\pm$0.05 &	0.14$\pm$0.04 &	0.28$\pm$0.05 \\			
\qquad Field   &   0.60$\pm$0.04 &	0.16$\pm$0.03 &	0.24$\pm$0.03 \\ 
\hline			
Red Spirals$^{\rm a}$ \\
\qquad Cluster & 0.70$\pm$0.04 &	0.25$\pm$0.04 &	0.05$\pm$0.02 \\			
\qquad Field   &   0.67$\pm$0.07 &	0.27$\pm$0.06 &	0.06$\pm$0.03 \\ 
\hline			
\end{tabular}
\end{center}
$^{\rm a}${Red spiral galaxies are discussed in \S\ref{sec:Continuo}.}
\end{table*}

\subsection{Sample}
\label{subsec:Muestra}

In the current work we use the sample presented in \citet{2012MNRAS.419.2703H}, which was employed there
to develop a structural diagnostic calibrated to detect mergers. This is a stellar mass and magnitude selected sample
($9.0 \leq \log M/M_{\odot}$; $R \leq 23.5\mathrm{mag}$) in the 
$0.05 \leq z_{\mathrm{phot}} \leq 0.30$ photometric redshift interval, and includes objects
in both the cluster and field environments of all morphological types, from elliptical to irregular galaxies.
The mass limit ensures that the sample is complete in stellar mass for both the blue cloud and the 
red sequence \citep[see][]{2006A&A...453..869B}. The magnitude limit guarantees reliable visual morphologies, since all the sources show extended
images in the HST/ACS data. The $R \leq 23.5\,\mathrm{mag}$ limit used to define the sample approximately corresponds to $m_{\rm F606W,AB}\simeq 24.0$ and therefore the structural parameters are reliable
for all members of this set of galaxies. The sample includes 1560 galaxies distributed across all environments.
The sample can also be divided into four different morphological 
categories labelled as ``E'', ``S0'', ``Sp'', and ``Irr''. The ``E'' bin is made of course of elliptical 
systems, the ``S0'' bin gathers the lenticular galaxies, the ``Sp'' bin comprises the spiral 
galaxies, and the ``Irr'' bin includes a mixture of irregulars, compact, and highly disturbed sources
that do not fit into any of the other galaxy classes.
The cluster sample is defined using a $R$-band-dependent photometric-redshift
running window designed to be 90\% complete in the apparent magnitude interval considered.
The contamination rate by field galaxies in the cluster sample is 30\% at the
fainter end, and substantially smaller at brighter magnitudes ($<10$\% for $R<20$). 
The field sample includes galaxies in the redshift intervals 
$z_{\mathrm{phot}} = [0.05, 0.14]$ and $z_{\mathrm{phot}} = [0.22, 0.30]$.
For a more detailed definition and justification of this sample, see \citet{2009MNRAS.393.1275G}.

The final sample consists of 655 Field galaxies (100 E, 60 S0, 318 Sp, 177 Irr) and 905 
Cluster galaxies (192 E, 216 S0, 383 Sp, 114 Irr).
The \textsc{STAGES} morphological catalogue has good morphological information for all objects 
in this sample. For more details about this sample and a complete summary of the motivations for its
definition, see \citet{2012MNRAS.419.2703H}.

This sample is split according to (i) whether or not galaxies are visually classified as mergers by \citet{2012MNRAS.419.2703H}, (ii) whether or not galaxies appear to have disturbed or merger-like structural parameters according 
to the diagnostic described in the same paper, and (iii) whether or not galaxies have a red spiral 
SED according to \citet{2005A&A...443..435W}. Throughout this work, galaxies visually classified as non-mergers  
will be divided into two classes, ``disturbed'' and ``relaxed'', according to  whether they have merger-like structural properties 
\citep[as defined by][]{2012MNRAS.419.2703H} or not. Specifically, non-mergers above the solid boundary line in 
the lower panel of Figure~10 of \citet{2012MNRAS.419.2703H} are classed as ``disturbed'' and  those below as ``relaxed''.

Figure~\ref{fig:binaryMergers0} allows to visualize the main subsamples that are relevant for this
work in the framework of Figure~10 of \citet{2012MNRAS.419.2703H}. It 
presents the Residual Flux Fraction (\textit{RFF}) \textsl{vs.} the asymmetry of the residual image after subtracting a S\'ersic
model calculated in a well defined way ($A_{\mathrm{CAS}}(\mathrm{Res})$). The
S\'ersic fits and structural parameters were measured over the HST F606W images
observed by the \textsc{STAGES} team. The \textit{RFF} is a measure of the amount of light present or absent in the
residuals that cannot be accounted for by the photometric errors. The asymmetry of the residuals is the
CAS \citep{2003ApJS..147....1C} asymmetry calculated for the residual image. It is important to note here that
the usual CAS parameters are calculated over the original, direct image while the structural parameters
introduced in \citet{2012MNRAS.419.2703H} are obtained using the residual image. This makes these measurements
different from other non-parametric structural characterizations of
surface-brightness distributions.

Figure~\ref{fig:binaryMergers0} shows, for both the cluster and field environments the distribution
of objects of all morphological classes in the
\textit{RFF}--$A_{\mathrm{CAS}}(\mathrm{Res})$ structural parameter space. The black, solid line
is the optimal boundary derived in \citet{2012MNRAS.419.2703H} that best
separates visual mergers from the general population
of galaxies. Objects below this line are morphologically relaxed systems, well represented by a S\'ersic model and shown as small black squares.
Objects above this line are disturbed galaxies, with substantial deviations from a simple S\'ersic model and shown
in Figure~\ref{fig:binaryMergers0} as blue circles. The location of the visual mergers is given by the
red filled infinity symbols. It is seen that, as expected, most {visual} mergers are disturbed systems and that the contamination
by disturbed systems not {visually classified as} merging, is very high. {It is important to remember that, as 
discussed above, some of these may be minor and/or advanced mergers.}
We note, however, that very few {visually identified} mergers lie below the boundary making this
diagnostic an excellent negative test. The subsample definitions shown in this figure will be used for
the plots and discussions presented in \S \ref{sec:Binario} and \S \ref{sec:Continuo}.

Table~\ref{tab:Fracciones1} gives the fractions of relaxed, disturbed, and
visual mergers (where applicable) for each of the morphological types in 
the cluster and field environments. Early types are mostly relaxed systems, while many Sp and Irr galaxies are 
either disturbed or merging objects. Errors are estimated assuming a binomial
distribution in which $\Delta f =\sqrt{ f\times(1-f)/N}$, where $f$ is the
estimated fraction and $N$ is the total number of events used to obtain
that fraction. {These estimates only take into account the statistical random errors. However, the contamination of the cluster sample by field galaxies (and vice-versa) would add some systematic uncertainty to these numbers which is very difficult to estimate without extensive spectroscopy reaching the faintest objects. However, since the fractions of field and cluster galaxies in the different categories (relaxed, disturbed and visual mergers) are very similar, we feel that these systematic errors would have little effect on our conclusions.} The table also presents the fractions corresponding to red spirals, which will be discussed later in the paper.

{It may seem surprising that, as Figure~\ref{fig:binaryMergers0} and Table~\ref{tab:Fracciones1} show, there is a significant fraction of irregular galaxies which are ``undisturbed'' according to the \textit{RFF}--$A_{\mathrm{CAS}}(\mathrm{Res})$ diagnostic. It is important to note, however, that the morphological classes (E, Sp, S0 and Irr) are determined visually, without reference to any quantitative structural parameter, while the \textit{RFF}--$A_{\mathrm{CAS}}(\mathrm{Res})$ diagnostic is based on a specific quantitative recipe which uses an empirically-calibrated threshold (cf. Fig.~\ref{fig:binaryMergers0} and \citealt{2012MNRAS.419.2703H} ). Galaxies above this threshold are deemed to be disturbed, while galaxies below are classified as relaxed. Although the fraction of irregular galaxies that are classified as ``undisturbed'' is, understandably, the smallest among all the morphologies, not all of them are disturbed enough to place them above the threshold. They simply are less disturbed  than the galaxies selected in \cite{2012MNRAS.419.2703H} as likely to be undergoing a merger.}

\section{The link between star formation history and structure}
\label{sec:Binario}

This section explores the differences in the star formation histories of galaxies 
classified according to their structural properties. In section \S\ref{sec:Continuo} 
we will expand the analysis of the star formation properties of galaxies as a continuous
function of their degree of morphological disturbance.

\subsection{Current star-formation rate}
\label{sec:SFR}

\begin{figure*}
\centering{
\subfigure{
\centering{\includegraphics[scale=0.45]{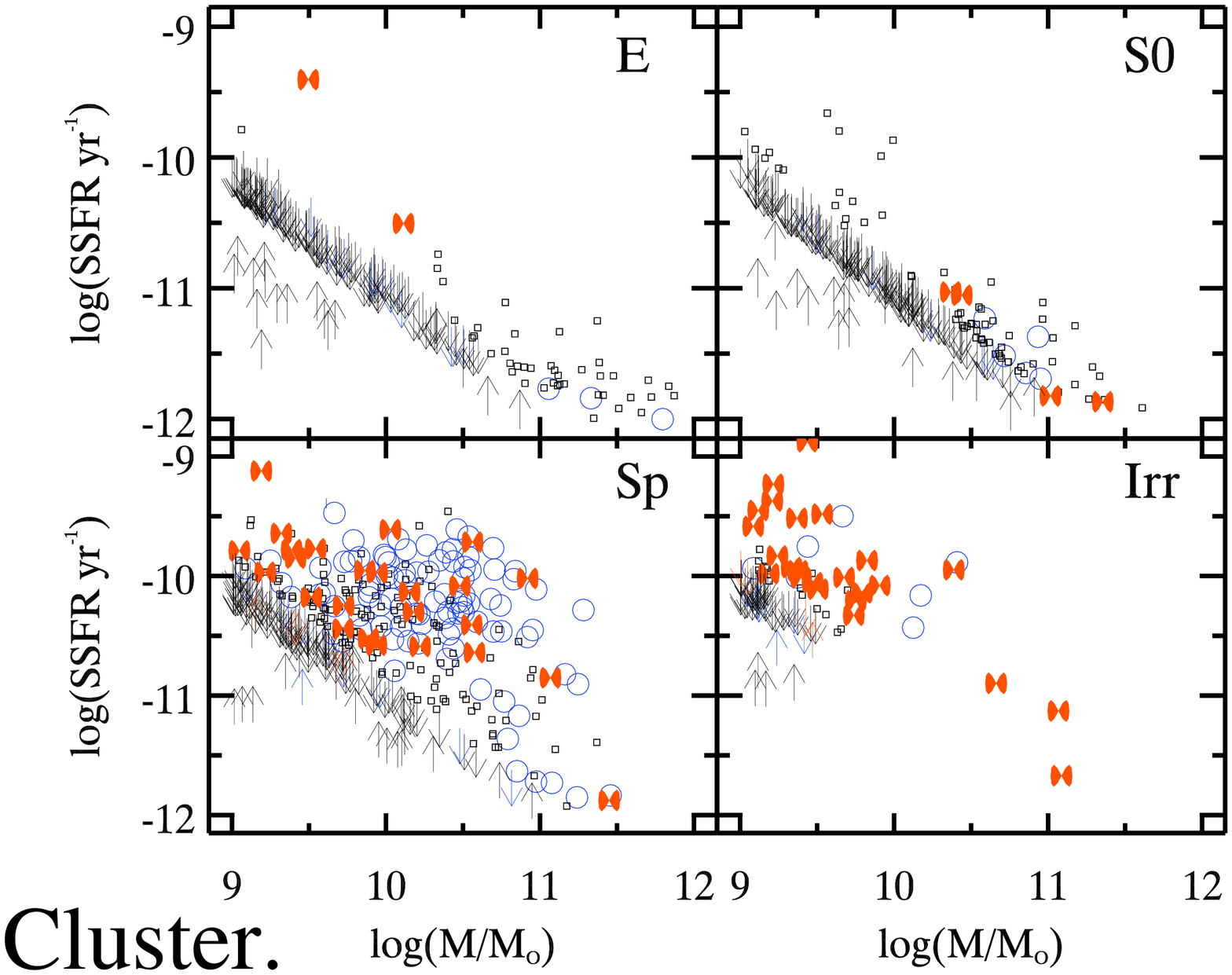}}}
\subfigure{
\centering{\includegraphics[scale=0.45]{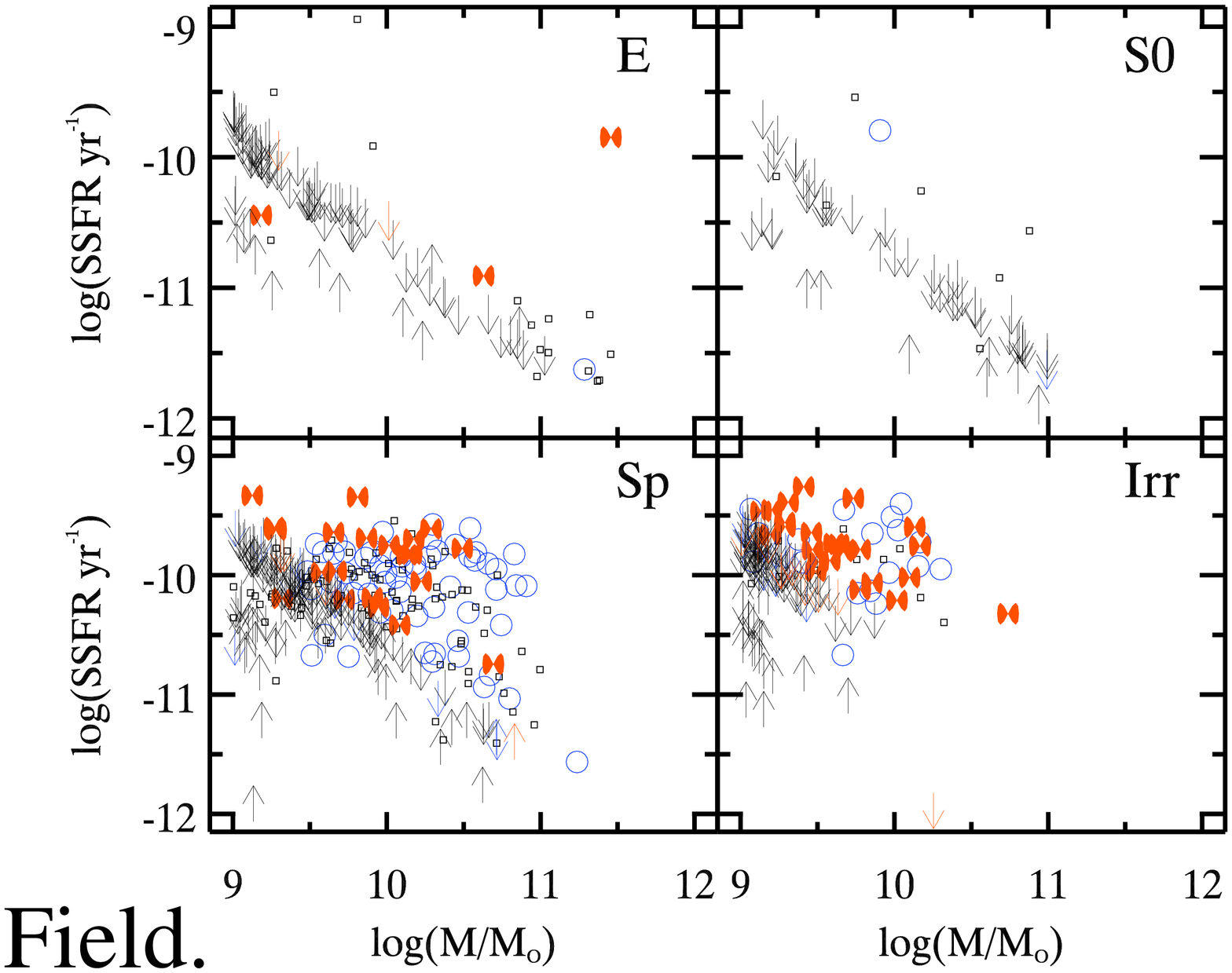}}}
\caption{\emph{Specific} SFR as a function of stellar mass for cluster (top panel) and field (bottom panel) galaxies 
separated by morphology. Relaxed systems are shown as small black
hollow squares ({\tiny$\square$}), disturbed objects {not visually classified as mergers} as blue
circles ($\bigcirc$), and visually classified mergers as red filled
infinity symbols ($\infty$). 
Objects for which only SFR upper limits are available  
are shown as down-pointing arrows with the  appropriate colour (black for 
relaxed galaxies, blue for disturbed non-merging galaxies, 
and red for visually-classified mergers).
In the same way, lower limits to the SFR measurements are shown as upward pointing arrows of the appropriate colour.
See text for details. 
Note that most arrows are black, meaning that most objects without
reliable measurements of the SFR are relaxed.
\label{fig:binaryMergers2}}
}
\end{figure*}
\begin{figure*}
\centering{
\centering{\includegraphics[scale=0.36]{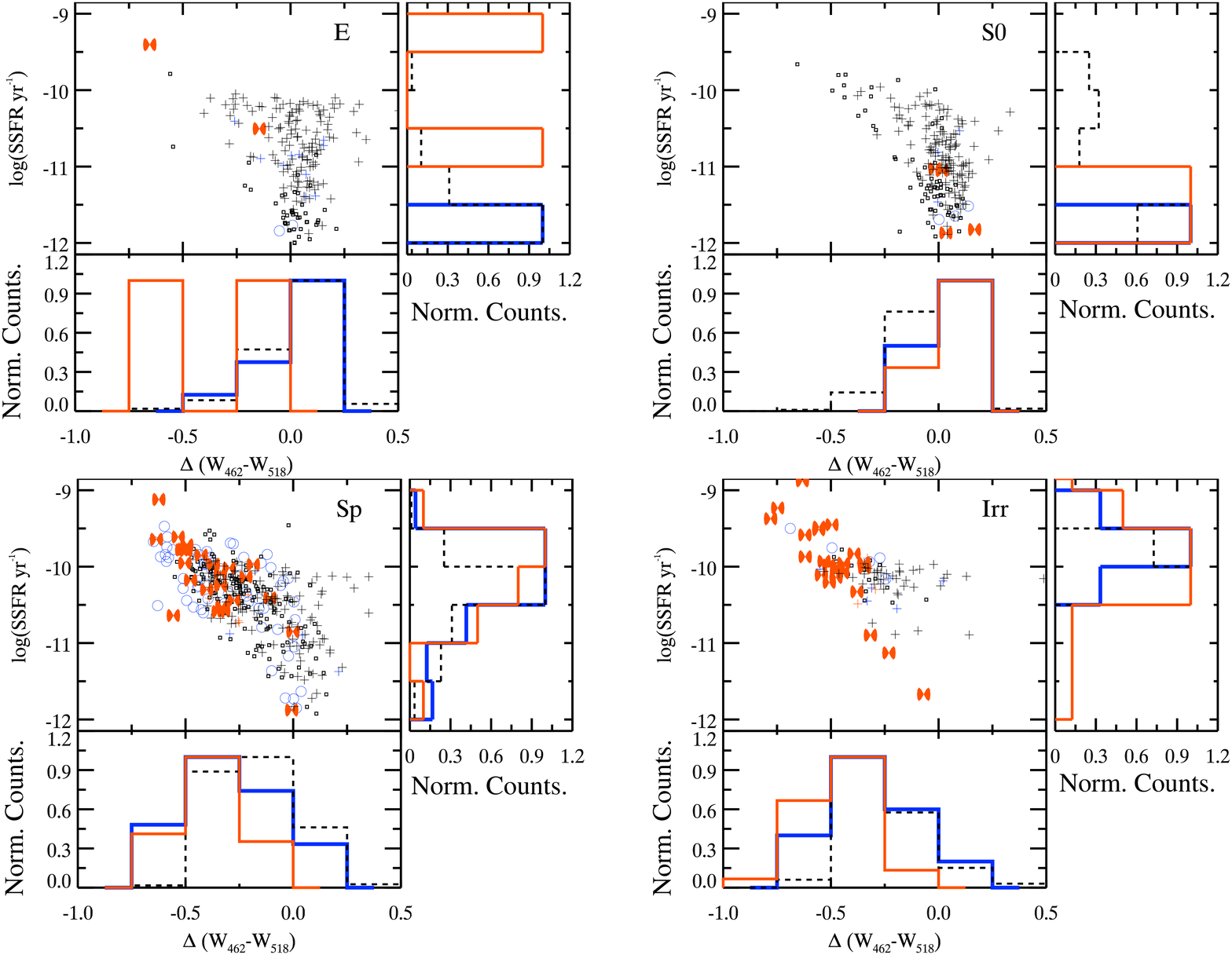}}
\caption{\emph{Specific} SFR \textsl{vs.} $\Delta (W_{462}-W_{518})$ for cluster galaxies of all morphologies. 
As in Fig.~\ref{fig:binaryMergers2}, black squares ({\tiny$\square$}) correspond to relaxed systems,
blue circles ($\bigcirc$)  to {disturbed galaxies not classified as mergers} and red ($\infty$) symbols to visual mergers.
Objects with only SFR upper or lower limits are presented as pluses ($+$) with appropriate colours.
In the accompanying histograms, the
black dashed line shows the relaxed galaxies, the blue solid line the disturbed ones, and the red histogram
the visual mergers. The SSFR histogram excludes objects with only SFR upper or lower limits. The $\Delta (W_{462}-W_{518})$
histogram includes all objects. In all cases, the histograms are normalized to a maximum value of one in order to amplify the differences between the
distribution functions.
\label{fig:binaryMergers3}}
}
\end{figure*}

\begin{figure*}
\centering{
\centering{\includegraphics[scale=0.36]{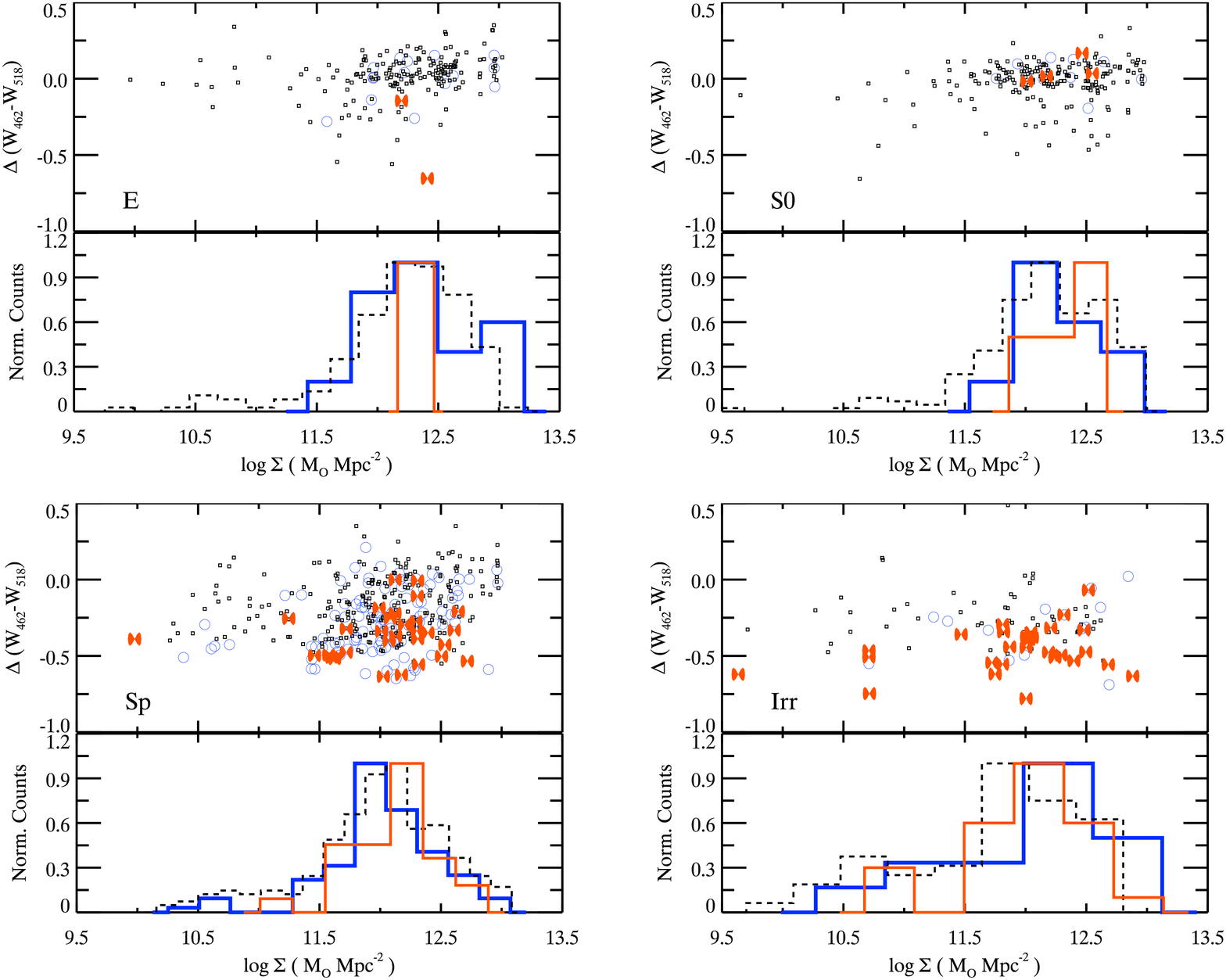}}
\caption{$\Delta (W_{462}-W_{518})$  \textsl{vs.} stellar mass density {$\Sigma$} (upper plot in each panel)
and stellar mass density histograms (lower plot in each panel) for the four different morphological types E, S0, Sp, Irr at
the cluster redshift. Symbols are as in Fig. \ref{fig:binaryMergers0}. Red,
filled infinity symbols ($\infty$) highlight the visual
mergers. The black-dashed histogram corresponds to relaxed systems, the blue thick line to {disturbed galaxies not classified as mergers}, and
the red line to visual mergers. 
Disturbed and visual mergers systems do not populate the lower density environments 
($\log \Sigma(\mathrm{M}_{\odot}\mathrm{Mpc}^{-2})\leq 11.5$, corresponding to
filaments and the in-fall region). 
At higher projected mass densities, all galaxies appear to be well mixed regardless of their morphological type
or degree of structural disturbance. Note that visual mergers are present even in the highest density environments
($\log \Sigma(\mathrm{M}_{\odot}\mathrm{Mpc}^{-2})\geq 12.5$), the cluster core. 
\label{fig:binaryMergers4}
}}
\end{figure*}

Figure~\ref{fig:binaryMergers2} presents the \emph{Specific} SFR (SSFR; star formation rate per unit stellar mass) 
as a function of the photometric stellar mass ($M_{\ast}/M_{\odot}$) for {disturbed galaxies not classified as 
mergers (blue circles), relaxed galaxies (small black squares), and visual mergers (red infinity symbols) in 
both the cluster and the field for all morphological types.}
The objects for which only SFR upper limits are available (cf. section~\ref{subsec:Datos})
are shown as down-pointing arrows with the  appropriate colour (black for 
relaxed galaxies, blue for {disturbed galaxies not classified as mergers}, 
and red for visually-classified mergers). Conversely, for the sources without
$24\micron$ \textit{Spitzer} \emph{observations}, their SFR are lower limits and are shown as
upward pointing arrows.

The clear diagonal pseudo-sequence defined by field and cluster Es and S0s, which consists mostly of upper limits and a few lower limits, 
shows the lowest total SFR that 
can be reliably measured with the \textsc{COMBO-17} and \textit{Spitzer} data
for the galaxies studied here.
Although the \textsc{COMBO-17} observations could in principle detect galaxies
up to two magnitudes fainter than our $R=23.5$ cut, 
the contamination resulting from the use of a much fainter limit would render the cluster subsample
unusable. However, even though lower SFRs, if present, could easily be detected by the deeper photometry,
an apparent lower limit in SFR results from the way we estimate it. 
The SFR is measured in part via a set of \textsc{PEGASE} models whose
specific assumptions need not be true for early type galaxies. One well-known deficiency of these models is
that the UV upturn \citep{1980ApJ...237L..65B,2011MNRAS.414.1887B}, which is not expected to be due to current star formation,
is not correctly accounted for. The 1300\AA\--2200\AA\ UV flux of old, massive
early type galaxies is often significant even in the total absence of star
formation. The effects of this extra flux can be measured in the
2600\AA\--3000\AA\ range covered by the 2800\AA\ (400\AA\ FWHM) synthetic filter
used to obtain the monochromatic UV luminosities.
This implies that a direct translation of UV luminosity into a SFR would be incorrect in these cases. Hence, this UV flux yields 
a non-zero SFR for these galaxies even if the true SFR is zero. 
Furthermore, the $24\micron$ flux, which is additionally 
used to obtain the SFR, is generally very low for early type galaxies when compared to spirals
and irregulars. Thus, with the flux density sensitivity of the \textit{Spitzer} observations ($58.0 \mu\mathrm{Jy}$), 
many E and S0 systems are not detected.  As explained in \S\ref{subsec:Datos}, these non-detections are 
still allocated some nominal SFR based on the IR detection upper limit, resulting in the appearance of 
an artificial minimum SFR threshold when none really exists. Hence, the
diagonal pseudo-sequence present in Fig.~\ref{fig:binaryMergers2} for E and S0
galaxies, consisting mostly of upper limits, is clearly due to selection/model effects.

The small offset of the field non-star-forming diagonal pseudo-sequence with respect to
that of the cluster sample is due to the slightly higher average redshift of the
field sample. Also, a number of spirals in the cluster non-star-forming pseudo
sequence are good candidates for anemic spirals or objects with large bulges, since the
\textsc{COMBO-17} colours mainly probe the inner regions of the galaxies as
explained in \S\ref{subsec:Datos}.

Figs.~\ref{fig:binaryMergers0} and~\ref{fig:binaryMergers2} clearly show that the majority of the E and S0
galaxies are morphologically smooth objects. More than 90\% of these galaxies
are classified as relaxed by the structural disturbance diagnostic, as is seen in Table~\ref{tab:Fracciones1}.
It is also seen that the very few E or S0 systems that appear to be disturbed or are visually classified as mergers
seem to follow the same general trends as the relaxed objects without any
clear mass segregation. However, a very small number of outliers do exist such as a couple of merging field E galaxies. In addition, the 
proportion of disturbed early-type galaxies appears to be 
around three times larger in the cluster than in the field. In fact, the vast majority of 
the disturbed E and S0 galaxies in this sample
are members of the A901/902 cluster complex. However, the 
small numbers involved make this result very uncertain. 
Nevertheless, it is clear that most early-type galaxies, and virtually all the ones in the field, seem to be 
systems which have not only completed their morphological transformation but
have passively-evolving stellar populations as well. {There is, however, a significant minority of S0s with relatively high SSFRs (see Fig.~\ref{fig:binaryMergers2}). This suggests that the process that suppresses star formation in these galaxies has not finished yet. If S0s are the descendants of spiral galaxies whose star formation ceased (see, e.g., \citealt{2006MNRAS.373.1125B,2012MNRAS.422.2590J}), star-forming S0s are those that have not been completely quenched yet, perhaps retaining some star formation in their central regions \citep{2012MNRAS.422.2590J}. }

The behaviour of the Sp and Irr sources in Fig.~\ref{fig:binaryMergers2} 
is fundamentally different from that of the E and S0 galaxies. The weakly star-forming pseudo-sequence is much less evident, indicating
that the SFR measurements for these sources are much more reliable. The weak-star-formation 
region of the diagram is usually populated by relaxed sources. Interestingly, the distributions in the 
the SSFR \textsl{vs.} $\log M_{\ast}/M_{\odot}$ diagram 
of visual mergers and {disturbed galaxies not classified as mergers} are indistinguishable.
These two classes of objects do not follow separate sequences, and
this is true for galaxies in the full mass range considered here. {There is no \textit{a-priori} reason why this had to be so: the physics driving star formation in mergers and non-mergers might not be the same, and different physics could lead to different behaviours on this diagram. However, that is not what we observe: their behaviours are indistinguishable. The mechanisms that lead to an enhancement in the SFR of disturbed and merging galaxies seem to ignore the origin of the structural properties of the systems in which they operate: 
a morphological disturbance is usually associated with an increased star
formation rate regardless of the mechanism that created such disturbance.
We also note that there are no clear differences between the cluster and field subsamples. }

{However, the current data do not allow us to rule out an alternative 
explanation, i.e., similar enhancements in the star formation could perturb 
the surface brightness distributions of galaxies to a comparable degree 
regardless of the origin of this increased star formation. In other words, it 
is not possible to ascertain whether the star formation is the
cause of the disturbance in the galaxies' surface brightness distribution, 
or, alternatively, a structural perturbation may be leading to an enhancement in the SFR. 
Regrettably, we are unable to clearly disentangle cause and effect here. 
It is arguable, though, that ---at least for major mergers--- the structural disturbance could come first and then the 
increase in SF happens. Mergers create asymmetries in the structure of the galaxies and also cause gas density enhancements. These may drive star-formation enhancements, manifesting themselves as additional structural disturbances in the optical images. In this case, the disturbed morphology measured in the images could be due both to the star-forming regions and to the initial asymmetries caused by the mergers.}

Figure \ref{fig:binaryMergers2} shows that, interestingly, the highest SSFR for the Sp and Irr systems in both the cluster
and field environments are attained in {visually identified} mergers. In this context, \citet{2009ApJ...704..324R} concluded  
that, even though the most intense star formation episodes taking place in massive 
($M>10^{10}M_{\odot}$) field galaxies at intermediate ($0.4<z<0.8$) redshifts are found in major mergers, 
the total impact of such interactions on the SFR amounts to only 10\% when averaged over the whole lifetimes of these systems.
Moreover, \citet{2009ApJ...697.1971J} found that massive ($M>2.5\times10^{10}M_{\odot}$) visual mergers
only contribute $\sim30\%$ to the total SFR density over comparable redshifts. 
Consistently, we find that {major} mergers in massive galaxies do not drive an enhancement 
in their SFR far beyond the one found in other disturbed galaxies. In other words, {major} mergers are
not the dominant trigger of star formation in galaxies at $z\leq 1.0$.

Although by their very nature irregulars contain the largest fraction of visual mergers, 
we emphasise that these galaxies show, fundamentally, a very similar behaviour to spirals in 
Figure~\ref{fig:binaryMergers2}. Disturbed galaxies have enhanced star formation whether they are {visually} classified as
mergers or not. This common behaviour is probably due to the fact that spirals and irregulars are 
generally gas rich, and enhanced star formation will be linked to a large gas density regardless of the actual mechanism
responsible for that large density. 

It could be argued that the statement ``disturbed galaxies have high SF irrespective of whether they are visually classified as mergers or not'' contains a circular argument driven solely by the fact that visual classification alone cannot unambiguously distinguish between mergers and non-mergers. However, we do not think that to be the case. Even though the the visual classification process cannot be completely objective and unequivocal, particularly in the absence of kinematic data, the classifiers' experience and expertise, informed by the examination of images of mergers generated in simulations, cannot be completely discounted. We are able to identify visual mergers as a distinct class of objects separate from the disturbed galaxies not classified as mergers. It is true that the latter class may contain minor and advanced mergers, but that does not mean that the visual merger class is not, at least statistically, distinct.  As discussed Section~\ref{sec:Background} and Fig.~\ref{fig:binaryMergers0} clearly shows, asymmetry or ``roughness'' are not enough to classify a galaxy as a visual merger. Additional diagnostics, such as the presence of tidal tails, double/multiple nuclei, clear bridges of material connecting different galaxy components, etcetera, are required. Thus, since the visual merger identification does contain information that is largely independent from the quantitative disturbance diagnostics, the above statement is not circular, although it is true that we can only be reasonably sure that it applies mostly to major mergers. Thus, the visual merger identification does contain information that is largely independent from the quantitative diagnostic of structural disturbance.

\subsection{Age of the stellar population}
\label{sec:SPAge}

The evidence presented so far indicates that the current star-formation properties of 
mergers and disturbed systems that are however {not classified as visual mergers}
are more similar than naively anticipated. At the same time, these findings
imply that disturbed surface brightness profiles can have several origins,
making it clear that such perturbations can be caused by either a merger or strong interaction,
or an internal event, {such as bar instabilities or shocks}, that can alter the surface brightness profile in a
similar manner. We now explore whether this similarity 
also applies to their past star-formation history.

Figure~\ref{fig:binaryMergers3} plots SSFR \textsl{vs.} $\Delta(W_{462}-W_{518})$.
The diagnostic $\Delta(W_{462}-W_{518})$ measures the offset of the \textsc{COMBO-17} medium-band colour 
$(W_{462}-W_{518})$ from a fit to the E and S0 red sequence, which was
calibrated using the reddening-corrected \textsc{COMBO-17} aperture-matched fluxes.
Explicitly, 
\begin{equation}
\Delta (W_{462}-W_{518}) = (W_{462}-W_{518}) - 2.551 + 0.0828 \times W_{518}.
\end{equation}
{The scatter of the E and S0 galaxies about this relation is quite small, ranging from $0.11$ at bright magnitudes to $0.18$ at the faint end ($0.15$ on average). Fig.~\ref{fig:binaryMergers3} gives an indication of the $\Delta(W_{462}-W_{518})$ range and scatter for the different morphologies.  }  

For cluster galaxies $(W_{462}-W_{518})$ measures the 4000\AA\ break, 
which is sensitive to stellar population age (see Fig.~6 in \citealt{2005A&A...443..435W}).
First-order age and metallicity effects are accounted for by subtracting off the  
mean colour-magnitude relation for passive galaxies (i.e., the red sequence). 
The reason behind this approach is that the red sequence for early type galaxies
indicates the locus of passive galaxies with different ages and metallicities,
as the chemical evolution is mainly driven here by astration. Thus, offsets blueward from this sequence are due to
the presence of younger stellar populations and $\Delta(W_{462}-W_{518})$ can be used as a reasonable proxy for 
the luminosity-weighted age of stellar populations on timescales of $\approx 10^8$--$10^{9} \mathrm{yr}$ 
\citep{2009MNRAS.393.1302W}.
Figure \ref{fig:binaryMergers3} also shows SSFR and $\Delta (W_{462}-W_{518})$ histograms
split according to the structural properties of the galaxies. All these
histograms are normalized to a maximum value of one.

E and S0 galaxies are predominantly on the red sequence. Since, {in most cases}, 
their  estimated SFRs can only be taken as upper limits  (as argued above), any
possible correlation would be masked. It is therefore not possible to say much on E and S0 galaxies from this figure.
However, notwithstanding the small fraction of merging and disturbed
early-type galaxies, there is no \emph{apparent} difference in the
distribution of merging and disturbed systems for the early-type galaxies on
this diagram. It is however stressed that small number statistics prevent us from reaching any meaningful
conclusion here.
{No striking difference are found between the full distributions of Es and S0s either, 
although there are a few S0s with significant SSFRs, as discussed in Section~\ref{sec:SFR}.} It
is important to point out, however, that given the narrow colour range and the
unreliable SFRs of these galaxies, these hints need to be taken with caution.

The SSFR correlates very strongly with $\Delta(W_{462}-W_{518})$ 
for the star-forming Sp and Irr galaxies. This is, perhaps, not surprising since 
both quantities are sensitive to star formation history (SFH), albeit on different timescales ($\sim 100\,$Myr vs. 
$\sim1\,$Gyr). {This difference in timescale sensitivity implies that, even if we expect most galaxies to follow this broad correlation, their exact position on this plane 
is not only dictated by the average age of their stellar populations but also by  
the ratio of current to average-past SFR. For instance, a galaxy for which this ratio is very high will be close to the upper envelope of the relation, while a galaxy with a low ratio will be close to the lower envelope.}

We find that the SSFR and $\Delta(W_{462}-W_{518})$ distributions for relaxed and disturbed galaxies 
are very different (compare black-dashed and blue histograms). 
In the case of the Sp galaxies, these differences can be summarized in the
lack of relaxed galaxies with $\Delta(W_{462}-W_{518})$ bluer than $-0.5$ and
stellar mass doubling times less than $10^{10}\mathrm{yr}$. These differences
translate into Kolmogorov-Smirnov probabilities of $2\times 10^{-5}$ and 0.019 for the
$\Delta(W_{462}-W_{518})$ and SSFR histograms respectively.
In the case of the Irr systems, it can also be seen that there are no relaxed
cluster Irr bluer than  $\Delta(W_{462}-W_{518})=-0.5$, and the SSFR limit is
in this case $10^{-9.5}\mathrm{yr}^{-1}$, which is a little bit more extreme.
However, the interpretation of the Kolmogorov-Smirnov probabilities is more
difficult here because of the smaller sample size ($\simeq 10$ \textsl{vs.}
$\simeq 60$ elements).
The SSFR and $\Delta(W_{462}-W_{518})$ histograms for the E and S0 galaxies
also show clear differences, although the differences are more obvious for the spirals due to better statistics.

Interestingly, the SSFR distributions
for visual mergers and {disturbed galaxies not classified as mergers} 
are indistinguishable from each other:
the red and blue SSFR histograms of the
Sp and Irr panels in Fig.~\ref{fig:binaryMergers3} have very similar shapes. 
Kolmogorov-Smirnov (K-S) tests were run in order to confirm this statement.
In the case of the Sp systems, the Kolmogorov-Smirnov probabilities are 0.055 and 0.48 for the
$\Delta(W_{462}-W_{518})$ and SSFR histograms respectively, and these same
probabilities are 0.05 and 0.94 for the Irr galaxies but again with a much
smaller sample size.
Moreover, the SSFR--$\Delta(W_{462}-W_{518})$ correlation has the same slope
for all galaxies (relaxed, disturbed non-mergers, and visual mergers).
{Since the location of a galaxy on the SSFR vs. $\Delta(W_{462}-W_{518})$ diagram 
is influenced by its SFH, and the distributions of the visual mergers and {the disturbed galaxies not  classified as mergers} 
are very similar, we can conclude that their SFHs should also be reasonably similar. At the very least, 
we can be certain that the differences in SFH must be small enough not to affect their distributions on this 
diagram.  Our findings are therefore consistent with the hypothesis that the SFH 
is similar in all disturbed galaxies, whether the disturbance is (major) merger-induced
or not. 


\begin{figure*}
\centering{
\centering{\includegraphics[scale=0.36]{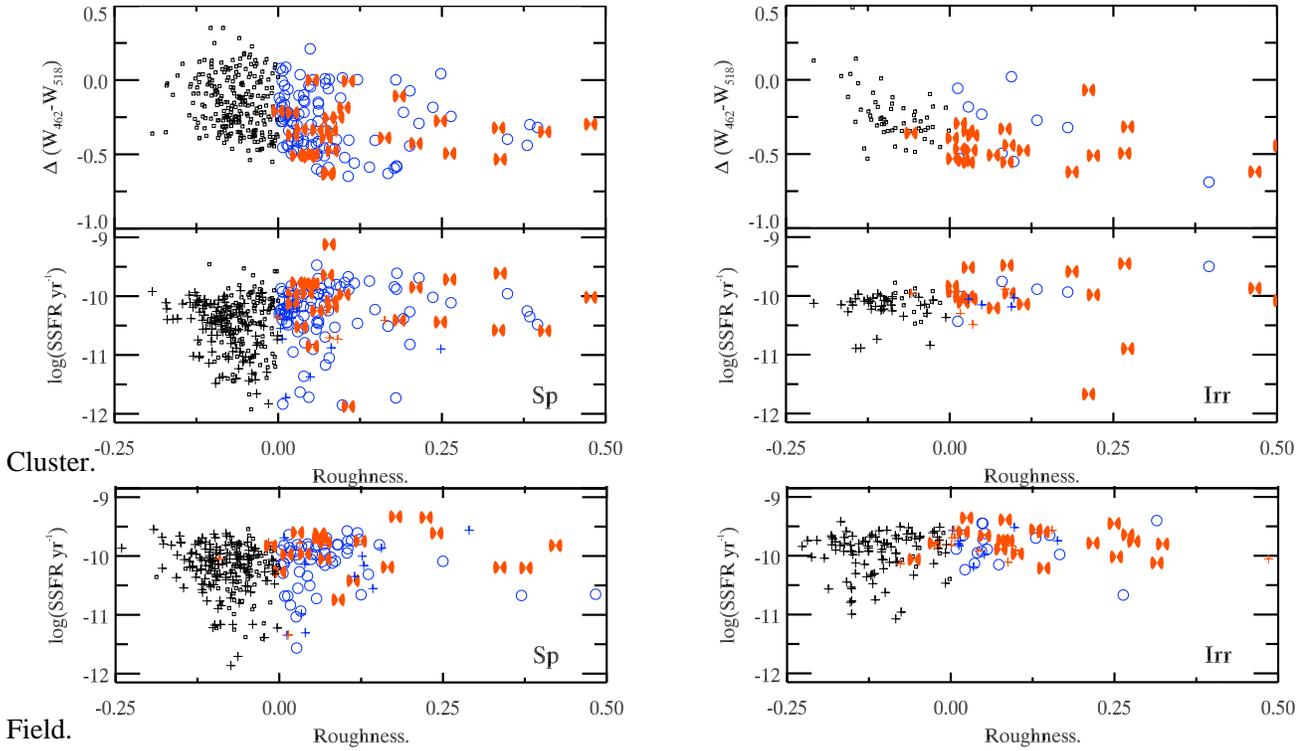}}
\caption{\emph{Specific} SFR and $\Delta(W_{462}-W_{518})$ {vs.} roughness
for cluster (top panels) and field (bottom panels) galaxies. 
$\Delta (W_{462}-W_{518})$ is only shown for cluster galaxies since it only works as an age indicator 
for galaxies at the cluster redshift (see text for details). Left panels show Sp systems whilst the right ones
present Irr galaxies. 
Symbols are as in Fig. \ref{fig:binaryMergers3}. 
For cluster galaxies, the mean age diagnostic
$\Delta(W_{462}-W_{518})$ is well correlated with the structural disturbance indicator for both Sp and Irr galaxies, regardless of whether 
they are {visual} mergers or not.  The correlation with the
instantaneous SSFR appears significantly weaker.
\label{fig:contMergers5}
}}
\end{figure*}

We close this section by exploring whether these results depend on environment within the A901/A902 system.  
Figure \ref{fig:binaryMergers4} presents $\Delta(W_{462}-W_{518})$ as a function
of the robust local environment measure $\log \Sigma (\mathrm{M}_{\odot}\mathrm{Mpc}^{-2})$ also
used in \citet[see \S\ref{subsec:Datos}]{2009MNRAS.393.1302W}.
The lack of an apparent correlation in the upper panels of
Fig.~\ref{fig:binaryMergers4} indicates that in the \textsc{STAGES} cluster system
it is possible to find galaxies with young and old underlying populations at
all local densities. {Figure~\ref{fig:binaryMergers4} also indicates that there is not a very strong
morphology--density relation in this multiple cluster system, suggesting that it is not yet completely relaxed.}

The lower panels in Fig.~\ref{fig:binaryMergers4} show histograms of
$\log \Sigma (\mathrm{M}_{\odot}\mathrm{Mpc}^{-2})$ for the four morphological
categories studied here, split according to their structural disturbance properties.
The histograms are normalized to a maximum value of one, and the bin width is
inversely proportional to the square root of the number of objects in each
histogram in order to ease the comparison.

It is interesting to note, first, that the low-density areas of the cluster system
($\log \Sigma(\mathrm{M}_{\odot}\mathrm{Mpc}^{-2})\leq 11.5$, corresponding to filaments and the in-fall region) 
are mainly populated by relaxed systems, in particular for the E and S0
morphological types where no disturbed or merging galaxies exist at these
densities. This is less clear for the Sp and Irr galaxies, and it is not
possible for a K-S test applied to the $\log \Sigma(\mathrm{M}_{\odot}\mathrm{Mpc}^{-2}) \leq 11.5$
data to give a definite answer for the Sp and Irr histograms because of the
small number of disturbed and visual merger galaxies in these regimes.
At higher densities, ($\log \Sigma(\mathrm{M}_{\odot}\mathrm{Mpc}^{-2}) > 11.5$), the three populations appear to be 
well mixed: a K-S test for the $\log \Sigma(\mathrm{M}_{\odot}\mathrm{Mpc}^{-2})\geq 11.5$ data fails to 
detect any significant differences between the distributions of the disturbed, relaxed, and visual mergers
of any morphology. The lowest K-S probability for each morphological category
is 0.64 between disturbed and relaxed E galaxies, 0.66 between relaxed and
merging S0 systems, 0.21 between relaxed and merging Sp galaxies and 0.20
between disturbed and relaxed Irr objects. The probabilities for all other
combinations in each morphological type are higher, and the results are more
significant for the Sp sources where the number counts are higher.
In addition, the Sp and Irr subpanels reveal that both disturbed non-mergers and visual mergers
with young stellar populations can be found even at the highest projected
densities ($\log \Sigma(\mathrm{M}_{\odot}\mathrm{Mpc}^{-2})\geq 12.5$), defined as the cluster ``core'' by 
\citet{2009MNRAS.393.1302W}.

The main conclusion of this section is that while there are differences
in the SSFR, colour and $\log \Sigma$ (at least when considering the full
density range) distributions of relaxed and disturbed galaxies, no such differences have been found {between the 
disturbed galaxies not classified as mergers and the visual mergers.}

\section{The link between star formation history and the degree of structural disturbance}
\label{sec:Continuo}

This section investigates whether there are any quantitative relations between the star-formation histories of galaxies (characterised 
by the $\mathrm{SSFR}$ and $\Delta (W_{462}-W_{518})$) and their degree of morphological
disturbance (parameterised using the merger diagnostic introduced by \citealt{2012MNRAS.419.2703H}).
This measurement, which we call ``roughness'' for brevity, is the perpendicular distance
to the boundary line defined in  the ``Residual Flux Fraction'' \textsl{vs.} ``Asymmetry of the Residuals''
diagram (see lower panel of Figure~10 in Hoyos et al.\ 2012 for details). 
This boundary is reproduced in Figure~\ref{fig:binaryMergers0} for completeness. 
Negative distances (corresponding to objects below the boundary) indicate relaxed systems while 
positive distances (objects above the boundary) imply disturbed ones. 

We reiterate here that positive roughness does not necessarily imply that the 
galaxy in question is a visual merger (cf. \S\ref{subsec:Muestra}): 
there is a very large fraction of disturbed objects
(as defined by the above criterion) that are not {visually classified as} mergers
\citep{2012MNRAS.419.2703H}. However, negative roughness does imply that a galaxy is 
\emph{not} a visual merger since virtually all the {visual mergers} are in the disturbed category.

\begin{figure*}
\centering{
\centering{\includegraphics[scale=0.45]{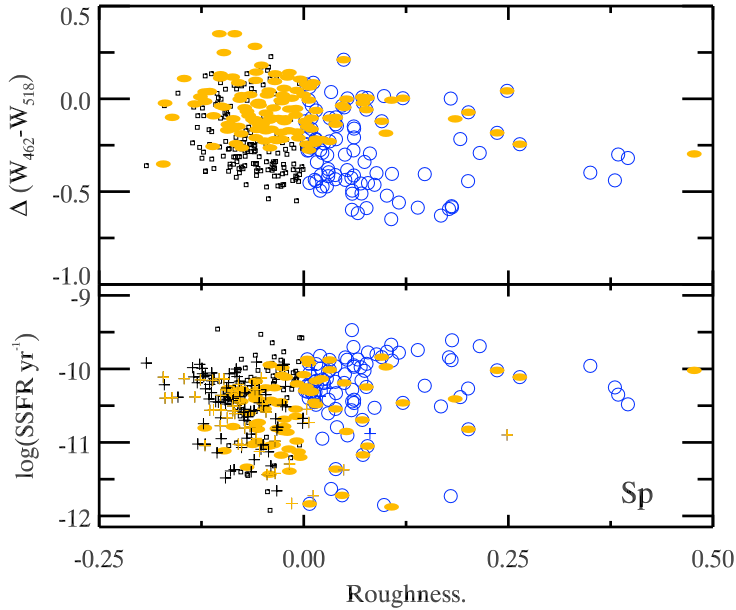}}
\caption{
\emph{Specific} SFR and $\Delta (W_{462}-W_{518})$ \textsl{vs.} roughness
for cluster spiral galaxies. This figure highlights the location of red spiral 
galaxies with reliable SFR as orange ellipses. 
As in Fig.~\ref{fig:binaryMergers3}, black hollow squares correspond to relaxed systems,
and blue circles to disturbed galaxies. In the lower panel, pluses ($+$) represent objects
with SFR upper or lower limits only.
It is very clear that passive spirals, which have been selected only through their integrated colours, 
{largely} have smooth profiles and thus their average roughness is negative ($-0.04\pm0.01$).
This is better seen in the upper panel, which is free of the representation problem related
with the determination of the SFR.
\label{fig:contRed5}
}}
\end{figure*}

Figure \ref{fig:contMergers5} plots  $\mathrm{SSFR}$ and $\Delta (W_{462}-W_{518})$
\textsl{vs.} roughness for cluster and field Sp and Irr galaxies. We do not show equivalent plots for E and S0 galaxies 
because most of these systems are not disturbed and their SSFRs are very low and often unreliable.   
Since $\Delta (W_{462}-W_{518})$ is explicitly designed to work as an age indicator for galaxies
at the cluster redshift (see \S\ref{sec:Binario}), we do not plot it for field galaxies.

There appears to be a good correlation between $\Delta (W_{462}-W_{518})$ (a proxy for stellar population age) 
and roughness: ``rougher'' objects have systematically 
younger stellar populations. A Spearman test 
indicates that this correlation is very significant for both cluster spirals ($\sim 7\sigma$) and irregulars 
($\sim 5\sigma$).  Furthermore, given 
the relatively bright and massive sample studied here, the correlation is not driven by luminosity or stellar mass, 
or affected by selection effects.  

Fig.~\ref{fig:contMergers5} also clearly shows that {visual} mergers and {disturbed galaxies not classified 
as mergers} have indistinguishable stellar age\footnote{This is also seen in Figure \ref{fig:binaryMergers3}.} and roughness distributions.
This reinforces our conclusion that the star formation histories of {disturbed galaxies not classified as mergers} 
and visual mergers appear to be {reasonably} similar. 
{It would seem as if the amount of ``roughness'' (or ``texture'' or ``structure'') 
determines the degree of star-formation enhancement in disturbed galaxies, \emph{regardless}
of whether a {(major)} merger is taking place. 
However, it is very difficult with the current data to separate cause
and effect so it is possible to argue that a similar degree of disturbance of
the surface brightness profile (roughness) are observed in galaxies with similar
enhancements of the SF and/or past star formation histories. Regardless, 
it is clear that ``roughness'' or ``sub-structure'' can be used as a proxy 
to distinguish galaxies with active star formation from those that 
do not have much on-going star-formation.
This fact is again consistent with the hypothesis  that the mechanisms 
that regulate the star formation events in galaxies act in the same manner whether 
the star formation is (major) merger-induced or not.}

Roughness appears to correlate less well with SSFR than with $\Delta (W_{462}-W_{518})$. 
The SSFR-roughness correlation is only statistically
significant for the cluster and field spirals at the $4\sigma$ level.
However, given the large uncertainties and scatter introduced in the determination of 
the galaxies' SFRs and stellar masses (see \S\ref{subsec:Datos}), 
it is likely that the underlying intrinsic correlation is stronger. 
Moreover, dust can have a stronger effect ---and thus introduce 
a larger scatter--- on the derived SFRs than 
on $\Delta (W_{462}-W_{518})$ given the narrow wavelength range spanned by this colour.
In any case, the highest SSFRs do correspond almost always to disturbed or merging
galaxies: very few relaxed cluster galaxies attain SSFRs greater
than $10^{-10}\mathrm{yr}^{-1}$.
It is therefore inadvisable to 
conclude that the strength of the correlations 
shown in Fig.~\ref{fig:contMergers5} is clearly different. What we can safely
conclude is that systems with similar degrees of roughness 
have a similar enhancement in their star formation rates on timescales 
$\sim 10^8$--$10^9\,$Gyrs, 
independently of the origin of  their morphological disturbances. 
{We must not forget, however, that there are many galaxies with relatively small roughness 
that have significant SSFR. Clearly, the existence of star formation linked 
with (or enhanced by) the structural disturbance does not preclude the existence of star formation smoothly 
distributed across the galaxies. }

To end this section we will approach the link between the degree of structural disturbance 
and star formation history from a different angle. We have found so far that galaxies with 
larger roughness have higher SSFRs and younger stellar ages.
We ask now whether the complementary statement is also true. 
Explicitly, if we select, independently of their structure, 
galaxies with suppressed recent star formation, do they exhibit low roughness? To answer this
question we consider the sample of optically-passive red spirals 
discussed in Sections~\ref{sec:Introduccion} and~\ref{subsec:Muestra}. These galaxies  
are known to have lower star formation rates than other spiral galaxies of similar masses \citep{2009MNRAS.393.1302W}. 
Figure~\ref{fig:contRed5} clearly shows that these galaxies,
selected solely from the properties of their SEDs \citep{2005A&A...443..435W}, have a
preference for smooth morphologies.
This is substantiated by the fact that the average roughness for
red cluster spirals is $-0.04\pm0.01$
whereas the average roughness for the other cluster spirals is
$-0.02\pm0.01$. This indicates that red spirals typically have 
slightly more negative roughness {(i.e., are smoother)} than other cluster spirals, although this is only a
two-sigma result.

Table~\ref{tab:Fracciones1} also shows that, in the cluster environment, 
red spiral galaxies have a slightly higher probability ($70 \pm 4$\%) of
showing a relaxed morphology than the global (i.e., red spirals plus other
cluster spiral galaxies) population ($65 \pm 2$\%). This difference is further
amplified if only the population of blue cloud cluster spirals, as defined by
their SED type in \citet{2005A&A...443..435W}, are considered. In this latter
case, only $63 \pm 2$\% are found to be relaxed.

This is in line with the findings presented in \citet{2013A&A...549A.142B}, which show that red spiral
galaxies are likely to be subject to enhanced ram pressure, when compared to normal spirals.
In that work, it is shown that morphologically undistorted galaxies show very strong asymmetries in their 
emission-line rotation curves at high rest-frame velocities. These kinematical asymmetries are thought to be caused by ram-pressure stripping
due to interaction with the intracluster medium. The work presented in \citet{2013A&A...549A.142B} also shows that, among disk galaxies, the
rotation-curve asymmetries are more acute in red spiral galaxies, leading to
think that ram-pressure stripping is an important factor in these systems,
probably leading to smoother morphologies through suppressed star formation.
The answer to the question above thus seems to be affirmative, and galaxies with
suppressed star formation show slightly smoother morphologies.

Finally, Table~\ref{tab:Fracciones1} also shows the fraction of relaxed, disturbed,
and visual mergers for red spiral galaxies. In this table, the total number of red spirals in
the cluster environment is more than double their number in the field, (122 \textsl{vs.} 48), highlighting that these red spirals are
mainly a cluster phenomenon. Nevertheless, red spirals are preferentially found among relaxed systems in both environments.

\section{Conclusions}
\label{sec:Conclusiones}

This paper presents a study of the links between star formation history and structure 
for a large mass-selected 
($\log M_{\ast}/M_{\odot}>9.0$) sample of morphologically-classified galaxies 
from the \textsc{STAGES} survey. 
More than half of these galaxies are located in the multiple cluster
system A901/902 ($z \approx 0.17$).
The remaining objects (the field sample) are made of both background ($0.22\leq z \leq 0.30$) and 
foreground ($0.05 \leq z \leq 0.14$) sources. 
The sample thus contains galaxies of all morphologies 
inhabiting a very broad range of environments,
from cluster cores to the field. 

We characterise the star-formation history of these galaxies by 
using specific star-formation rates 
and stellar-population age proxies derived from extensive
UV, optical and infrared photometry. The disturbance in the galaxies' structure is quantified using  
objective measurements based on HST images following \citet{2012MNRAS.419.2703H}. The sample is divided 
into undisturbed galaxies, {disturbed galaxies that are not visually classified as merging},  
and disturbed galaxies which are
visually {classified} as mergers. Moreover, a quantitative measurement of the degree of disturbance 
(which we call ``roughness'') is calculated for each galaxy. 

The main conclusions of this study are:

\begin{itemize}

\item{As expected, the vast majority of E and S0 galaxies in all environments  
have relaxed structure and show no signs of ongoing or recent star formation.}

\item{Structurally-disturbed galaxies have higher specific star-formation rates and 
younger stellar populations than their relaxed counterparts, and
they tend to avoid the lowest density regions.}

\item{Conversely, cluster spirals with reduced/quenched star
formation (red spirals) seem to have somewhat less disturbed (smoother) morphologies
than spirals with ``normal'' star-formation activity. This would fit into
the scenario that these ``passive'' spirals have started their
morphological transformation towards becoming lenticulars.}

\item{{Galaxies visually identified as merging and equally disturbed (but not classified as merging) ones have very similar specific star-formation rates and stellar ages. They also reside in very similar environments.}}

\end{itemize}

{The specific star formation rate in galaxies appears to be linked to the degree of 
structural disturbance (measured by the ``roughness'' of their images), 
regardless of its physical cause. Specifically, an increase in the galaxies' 
``roughness'' or ``texture'' (i.e., deviations from a smooth profile) will result 
in a proportional increase in their star formation, whether they are visually identified 
as merging or not. In this interpretation, we speculate that
merging galaxies are not special in terms of their higher-than-normal
star-formation activity. Any physical process that produces ``roughness'', or, in other words, regions
of enhanced luminosity density in a galaxy, will
increase the star-formation activity in the galaxy with similar efficiency. 
It is also possible to argue that galaxies with similar specific star
formation rates will exhibit similar ``roughness''. 
In this interpretation, it is not possible to tell what is
the physical origin of an enhancement in the SFR of a given galaxy using 
morphological diagnostics like the one used here. Further
data, possibly of kinematic nature, are needed to do this.
Nevertheless, ``roughness'' or ``sub-structure'' can be used as a proxy 
to distinguish galaxies with active star formation from those without.
}

\section*{Acknowledgments}

CH acknowledges funding from the Spanish AYA2010-21887-C04-03 Grant, led by Prof. \'Angeles D\'{\i}az. 
Support for STAGES was provided by NASA through GO-10395 from STScI operated by AURA under NAS5- 26555. 
AAS, MEG and DTM acknowledge financial support from the UK Science and Technology Facilities Council.
AB acknowledges funding by the Austrian Science Foundation FWF (projects
P19300-N16 and P23946-N16). We acknowledge the use of data from the following observing facilities:
{MPG/ESO 2.2-m}, {HST (ACS)} and {Spitzer (MIPS)}. We thank the anonymous referee for her/his constructive 
criticisms and comments which have significantly contributed to the improvement of the original manuscript.


\label{lastpage}

\begin{thebibliography}{}

\bibitem[Bergvall et al.(2003)]{2003A&A...405...31B}
Bergvall, N., Laurikainen, E., \& Aalto, S.\ 2003, \aap, 405, 31 

\bibitem[Barnes \& Hernquist(1991)]{1991ApJ...370L..65B}
Barnes, J.~E., \& Hernquist, L.~E.\ 1991, \apjl, 370, L65 

\bibitem[Barnes \& Hernquist(1996)]{1996ApJ...471..115B}
Barnes, J.~E., \& Hernquist, L.\ 1996, \apj, 471, 115 

\bibitem[Bedregal et al.(2006)]{2006MNRAS.373.1125B} Bedregal, A.~G., 
Arag{\'o}n-Salamanca, A., \& Merrifield, M.~R.\ 2006, \mnras, 373, 1125 

\bibitem[Bell et al.(2005)]{2005ApJ...625...23B} Bell, E.~F., Papovich, C., 
Wolf, C., et al.\ 2005, \apj, 625, 23 

\bibitem[Bell et al.(2004)]{2004ApJ...608..752B}
Bell, E.~F., Wolf, C., Meisenheimer, K., et al.\ 2004, \apj, 608, 752 

\bibitem[Bertola et al.(1980)]{1980ApJ...237L..65B}
Bertola, F., Capaccioli, M., Holm, A.~V., \& Oke, J.~B.\ 1980, \apjl, 237, L65 

\bibitem[Bluck et al.(2012)]{2012ApJ...747...34B}
Bluck, A.~F.~L., Conselice, C.~J., Buitrago, F., et al.\ 2012, \apj, 747, 34 

\bibitem[Borch et al.(2006)]{2006A&A...453..869B}
Borch, A., Meisenheimer, K., Bell, E.~F., et al.\ 2006, \aap, 453, 869 

\bibitem[Boselli \& Gavazzi(2006)]{2006PASP..118..517B} Boselli, A., \& Gavazzi, G.\ 2006, \pasp, 118, 517 

\bibitem[Bournaud et al.(2007)]{2007A&A...476.1179B}
Bournaud, F., Jog, C.~J., \& Combes, F.\ 2007, \aap, 476, 1179 

\bibitem[Bournaud et al.(2004)]{2004A&A...418L..27B}
Bournaud, F., Combes, F., \& Jog, C.~J.\ 2004, \aap, 418, L27 

\bibitem[Bournaud et al.(2005)]{2005A&A...437...69B}
Bournaud, F., Jog, C.~J., \& Combes, F.\ 2005, \aap, 437, 69 

\bibitem[B{\"o}sch et al.(2013)]{2013A&A...549A.142B}
B{\"o}sch, B., B{\"o}hm, A., Wolf, C., et al.\ 2013, \aap, 549, A142 

\bibitem[Bureau et al.(2011)]{2011MNRAS.414.1887B}
Bureau, M., Jeong, H., Yi, S.~K., et al.\ 2011, \mnras, 414, 1887 

\bibitem[Conselice(2003)]{2003ApJS..147....1C}
Conselice, C.~J.\ 2003, \apjs, 147, 1 

\bibitem[De Lucia \& Blaizot(2007)]{2007MNRAS.375....2D}
De Lucia, G., \& Blaizot, J.\ 2007, \mnras, 375, 2 

\bibitem[Di Matteo et al.(2005)]{2005Natur.433..604D} 
Di Matteo, T., Springel, V., \& Hernquist, L.\ 2005, \nat, 433, 604 

\bibitem[Di Matteo et al.(2008)]{2008A&A...492...31D}
Di Matteo, P., Bournaud, F., Martig, M., et al.\ 2008, \aap, 492, 31 

\bibitem[Dressler(1980)]{1980ApJ...236..351D}
Dressler, A.\ 1980, \apj, 236, 351 

\bibitem[Duc(2011)]{2011arXiv1101.4834D}
Duc, P.-A.\ 2011, arXiv:1101.4834 

\bibitem[Faber et al.(2007)]{2007ApJ...665..265F}
Faber, S.~M., Willmer, C.~N.~A., Wolf, C., et al.\ 2007, \apj, 665, 265 

\bibitem[Fioc \& Rocca-Volmerange(1997)]{1997A&A...326..950F}
Fioc, M., \& Rocca-Volmerange, B.\ 1997, \aap, 326, 950 

\bibitem[Fujita(1998)]{1998ApJ...509..587F}
Fujita, Y.\ 1998, \apj, 509, 587 

\bibitem[Gallazzi et al.(2009)]{2009ApJ...690.1883G}
Gallazzi, A., Bell, E.~F., Wolf, C., et al.\ 2009, \apj, 690, 1883 

\bibitem[Gonz{\'a}lez-Garc{\'{\i}}a \& Balcells(2005)]{2005MNRAS.357..753G}
Gonz{\'a}lez-Garc{\'{\i}}a, A.~C., \& Balcells, M.\ 2005, \mnras, 357, 753 

\bibitem[Gray et al.(2004)]{2004MNRAS.347L..73G}
Gray, M.~E., Wolf, C., Meisenheimer, K., et al.\ 2004, \mnras, 347, L73 

\bibitem[Gray et al.(2009)]{2009MNRAS.393.1275G}
Gray, M.~E., Wolf, C., Barden, M., et al.\ 2009, \mnras, 393, 1275 

\bibitem[Gray et al.(in prep)]{StagesMorfoCat}
Gray, M. E., et al. \textit{in prep.}

\bibitem[Gunn \& Gott(1972)]{1972ApJ...176....1G}
Gunn, J.~E., \& Gott, J.~R., III 1972, \apj, 176, 1 

\bibitem[Gy{\H o}ry \& Bell(2010)]{2010ApJ...724..694G}
Gy{\H o}ry, Z., \& Bell, E.~F.\ 2010, \apj, 724, 694 

\bibitem[Hashimoto et al.(1998)]{1998ApJ...499..589H} Hashimoto, Y., 
Oemler, A., Jr., Lin, H., \& Tucker, D.~L.\ 1998, \apj, 499, 589 

\bibitem[Heiderman et al.(2009)]{2009ApJ...705.1433H} Heiderman, A., Jogee, 
S., Marinova, I., et al.\ 2009, \apj, 705, 1433 

\bibitem[Hopkins et al.(2005)]{2005ApJ...630..705H}
Hopkins, P.~F., Hernquist, L., Cox, T.~J., et al.\ 2005, \apj, 630, 705 

\bibitem[Hoyos et al.(2004)]{2004AJ....128.1541H}
Hoyos, C., Guzm{\'a}n,  R., Bershady, M.~A., Koo, D.~C., \& D{\'{\i}}az, A.~I.\ 2004, \aj, 128, 1541 

\bibitem[Hoyos et al.(2012)]{2012MNRAS.419.2703H}
Hoyos, C., Arag{\'o}n-Salamanca, A., Gray, M.~E., et al.\ 2012, \mnras, 419, 2703 

\bibitem[Hunter(1997)]{1997PASP..109..937H}
Hunter, D.\ 1997, \pasp, 109, 937 

\bibitem[Jogee et al.(2009)]{2009ApJ...697.1971J}
Jogee, S., Miller, S.~H., Penner, K., et al.\ 2009, \apj, 697, 1971 

\bibitem[Johansson et al.(2009)]{2009ApJ...707L.184J} 
Johansson, P.~H., Burkert, A., \& Naab, T.\ 2009, \apjl, 707, L184 


\bibitem[Johnston et al.(2012)]{2012MNRAS.422.2590J} Johnston, E.~J., 
Arag{\'o}n-Salamanca, A., Merrifield, M.~R., 
\& Bedregal, A.~G.\ 2012, \mnras, 422, 2590 

\bibitem[Kennicutt(1998)]{1998ARA&A..36..189K}
Kennicutt, R.~C., Jr.\ 1998, \araa, 36, 189 

\bibitem[Koopmann \& Kenney(2004)]{2004ApJ...613..866K}
Koopmann, R.~A., \& Kenney, J.~D.~P.\ 2004, \apj, 613, 866 

\bibitem[Larson et al.(1980)]{1980ApJ...237..692L}
Larson, R.~B., Tinsley, B.~M., \& Caldwell, C.~N.\ 1980, \apj, 237, 692 

\bibitem[Lewis et al.(2002)]{2002MNRAS.334..673L} Lewis, I., Balogh, M., De 
Propris, R., et al.\ 2002, \mnras, 334, 673 

\bibitem[L{\'o}pez-Sanjuan et al.(2010a)]{2010ApJ...710.1170L} 
L{\'o}pez-Sanjuan, C., Balcells, M., P{\'e}rez-Gonz{\'a}lez, P.~G., et al.\ 2010a, \apj, 710, 1170 

\bibitem[L{\'o}pez-Sanjuan et al.(2010b)]{2010A&A...518A..20L}
L{\'o}pez-Sanjuan, C., Balcells, M., P{\'e}rez-Gonz{\'a}lez, P.~G., et al.\ 2010b, \aap, 518, A20 

\bibitem[L{\'o}pez-Sanjuan et al.(2012)]{2012arXiv1202.4674L}
L{\'o}pez-Sanjuan, C., Le F{\`e}vre, O., Ilbert, O., et al.\ 2012, arXiv:1202.4674 

\bibitem[Lotz et al.(2004)]{2004AJ....128..163L} Lotz, J.~M., Primack, J., 
\& Madau, P.\ 2004, \aj, 128, 163 

\bibitem[Lotz et al.(2010)]{2010MNRAS.404..590L}
Lotz, J.~M., Jonsson, P., Cox, T.~J., \& Primack, J.~R.\ 2010, \mnras, 404, 590 

\bibitem[Melnick \& Sargent(1977)]{1977ApJ...215..401M}
Melnick, J., \& Sargent, W.~L.~W.\ 1977, \apj, 215, 401 

\bibitem[Mihos \& Hernquist(1996)]{1996ApJ...464..641M}
Mihos, J.~C., \& Hernquist, L.\ 1996, \apj, 464, 641 

\bibitem[Moore et al.(1998)]{1998ApJ...495..139M}
Moore, B., Lake, G., \& Katz, N.\ 1998, \apj, 495, 139 

\bibitem[Naab \& Trujillo(2006)]{2006MNRAS.369..625N}
Naab, T., \& Trujillo, I.\ 2006, \mnras, 369, 625 

\bibitem[Nipoti \& Binney(2007)]{2007MNRAS.382.1481N} Nipoti, C., \& Binney, J.\ 2007, \mnras, 382, 1481 

\bibitem[Patton et al.(2000)]{2000ApJ...536..153P} Patton, D.~R., Carlberg, 
R.~G., Marzke, R.~O., et al.\ 2000, \apj, 536, 153 

\bibitem[Poggianti et al.(1999)]{1999ApJ...518..576P}
Poggianti, B.~M., Smail, I., Dressler, A., et al.\ 1999, \apj, 518, 576 

\bibitem[Poggianti et al.(2008)]{2008ApJ...684..888P} Poggianti, B.~M., 
Desai, V., Finn, R., et al.\ 2008, \apj, 684, 888 

\bibitem[Poggianti et al.(2006)]{2006ApJ...642..188P} Poggianti, B.~M., von 
der Linden, A., De Lucia, G., et al.\ 2006, \apj, 642, 188 

\bibitem[Rix et al.(2004)]{2004ApJS..152..163R} 
Rix, H.-W., Barden, M., Beckwith, S.~V.~W., et al.\ 2004, \apjs, 152, 163 

\bibitem[Robaina et al.(2009)]{2009ApJ...704..324R} 
Robaina, A.~R., Bell, E.~F., Skelton, R.~E., et al.\ 2009, \apj, 704, 324 

\bibitem[Robaina et al.(2010)]{2010ApJ...719..844R}
Robaina, A.~R., Bell, E.~F., van der Wel, A., et al.\ 2010, \apj, 719, 844 

\bibitem[Schweizer(1982)]{1982ApJ...252..455S}
Schweizer, F.\ 1982, \apj, 252, 455 

\bibitem[Schweizer \& Seitzer(1992)]{1992AJ....104.1039S}
Schweizer, F., \& Seitzer, P.\ 1992, \aj, 104, 1039 

\bibitem[Schweizer(1999)]{1999Ap&SS.267..299S}
Schweizer, F.\ 1999, \apss, 267, 299 

\bibitem[Smith et al.(2005)]{2005ApJ...620...78S}
Smith, G.~P., Treu, T., Ellis, R.~S., Moran, S.~M., \& Dressler, A.\ 2005, \apj, 620, 78 

\bibitem[Spitzer \& Baade(1951)]{1951ApJ...113..413S}
Spitzer, L., Jr., \& Baade, W.\ 1951, \apj, 113, 413 

\bibitem[Stanford \& Bushouse(1991)]{1991ApJ...371...92S}
Stanford, S.~A., \& Bushouse, H.~A.\ 1991, \apj, 371, 92 

\bibitem[Toomre(1977)]{1977egsp.conf..401T}
Toomre, A.\ 1977, Evolution of Galaxies and Stellar Populations, 401 

\bibitem[Trujillo et al.(2011)]{2011MNRAS.415.3903T}
Trujillo, I., Ferreras, I., \& de La Rosa, I.~G.\ 2011, \mnras, 415, 3903 

\bibitem[van den Bergh(1976)]{1976ApJ...206..883V}
van den Bergh, S.\ 1976, \apj, 206, 883 

\bibitem[Wen et al.(2009)]{2009ApJ...692..511W}
Wen, Z.~L., Liu, F.~S., \& Han, J.~L.\ 2009, \apj, 692, 511 

\bibitem[Wolf et al.(2003)]{2003A&A...401...73W}
Wolf, C., Meisenheimer, K., Rix, H.-W., et al.\ 2003, \aap, 401, 73 

\bibitem[Wolf et al.(2004)]{2004A&A...421..913W} Wolf, C., Meisenheimer, K., Kleinheinrich, M., et al.\ 2004, \aap, 421, 913 

\bibitem[Wolf et al.(2005)]{2005ApJ...630..771W} Wolf, C., Bell, E.~F., 
McIntosh, D.~H., et al.\ 2005, \apj, 630, 771 

\bibitem[Wolf et al.(2005)]{2005A&A...443..435W}
Wolf, C., Gray, M.~E., \& Meisenheimer, K.\ 2005, \aap, 443, 435 

\bibitem[Wolf et 
al.(2008)]{2008A&A...492..933W} Wolf, C., Hildebrandt, H., Taylor, E.~N., \& Meisenheimer, K.\ 2008, \aap, 492, 933 

\bibitem[Wolf et al.(2009)]{2009MNRAS.393.1302W} 
Wolf, C., Arag{\'o}n-Salamanca, A., Balogh, M., et al.\ 2009, \mnras, 393, 1302 

\end{thebibliography}
\end{document}